\documentclass[runningheads]{llncs}
\usepackage{graphicx}
\usepackage{subcaption}
\usepackage{calc}
\usepackage{amssymb}
\usepackage{tabularx}
\usepackage[bottom]{footmisc}
\usepackage{tikz}
\usetikzlibrary{quantikz}
\usepackage{adjustbox}
\usepackage{hyperref}
\usepackage{comment}
\usepackage{booktabs}
\usepackage{todonotes}
\begin{document}
\title{Weight Re-Mapping for Variational Quantum Algorithms}
%
%
\author{Michael Kölle\inst{1} \and 
Alessandro Giovagnoli\inst{1} \and
Jonas Stein\inst{1} \and
Maximilian Balthasar Mansky\inst{1} \and
Julian Hager\inst{1} \and
Tobias Rohe\inst{1} \and
Robert Müller\inst{1} \and
Claudia Linnhoff-Popien\inst{1}}
\authorrunning{M. Kölle et al.}
%
\institute{LMU Munich, Oettingenstraße 67, 80538 Munich, Germany 
\email{\{michael.koelle\}@ifi.lmu.de}}
\maketitle              
\begin{abstract}
Inspired by the remarkable success of artificial neural networks across a broad spectrum of AI tasks, variational quantum circuits (VQCs) have recently seen an upsurge in quantum machine learning applications. The promising outcomes shown by VQCs, such as improved generalization and reduced parameter training requirements, are attributed to the robust algorithmic capabilities of quantum computing.
However, the current gradient-based training approaches for VQCs do not adequately accommodate the fact that trainable parameters (or weights) are typically used as angles in rotational gates. To address this, we extend the concept of weight re-mapping for VQCs, as introduced by Kölle et al. \cite{koelle2023}. This approach unambiguously maps the weights to an interval of length $2\pi$, mirroring data rescaling techniques in conventional machine learning that have proven to be highly beneficial in numerous scenarios.
In our study, we employ seven distinct weight re-mapping functions to assess their impact on eight classification datasets, using variational classifiers as a representative example. Our results indicate that weight re-mapping can enhance the convergence speed of the VQC. We assess the efficacy of various re-mapping functions across all datasets and measure their influence on the VQC's average performance. Our findings indicate that weight re-mapping not only consistently accelerates the convergence of VQCs, regardless of the specific re-mapping function employed, but also significantly increases accuracy in certain cases.
\end{abstract}

\section{\uppercase{Introduction}}
\label{sec:introduction}

In past decades, machine learning (ML) has become an indispensable tool for tackling a wide range of problems in both science and industry. Overcoming previously intractable computational problems, ML has nevertheless been faced with intrinsic limitations, such as the curse of dimensionality \cite{bellman1966dynamic}. The successes of quantum computing and machine learning have given rise to quantum machine learning (QML), now a central topic in both artificial intelligence and quantum computing.
QML is an approach to ML based on the laws of quantum mechanics. The basic unit of quantum information, the qubit, can be exploited to encode information and through quantum manipulations problems can be solved computationally faster than on a classical computer. In some cases, such techniques provide a quantum advantage, as seen in quantum algorithms used to solve systems of linear equations \cite{Harrow_2009}. Specifically, this topic often centers around the acceleration of basic linear algebra subroutines. 
A major focus in the field of QML is Variational Quantum Computing (VQC) which employs quantum gates. Considering a qubit as the basic unit of information---a 2-dimensional complex vector in a Hilbert space---quantum gates can be represented as complex unitary operators, rotating or reflecting the vector on a unitary sphere. Every quantum gate can be decomposed into rotations and reflections \cite{nielsen_chuang_2010}. Concatenating quantum gates gives rise to a quantum counterpart of the classical neural network: a quantum function approximator. The use of a qubit as a basic information unit allows for exponentially larger information encoding, as the space it occupies expands exponentially with the dimensions.\\

\begin{figure}[t]
    \centering
    \includegraphics[width=0.7\linewidth]{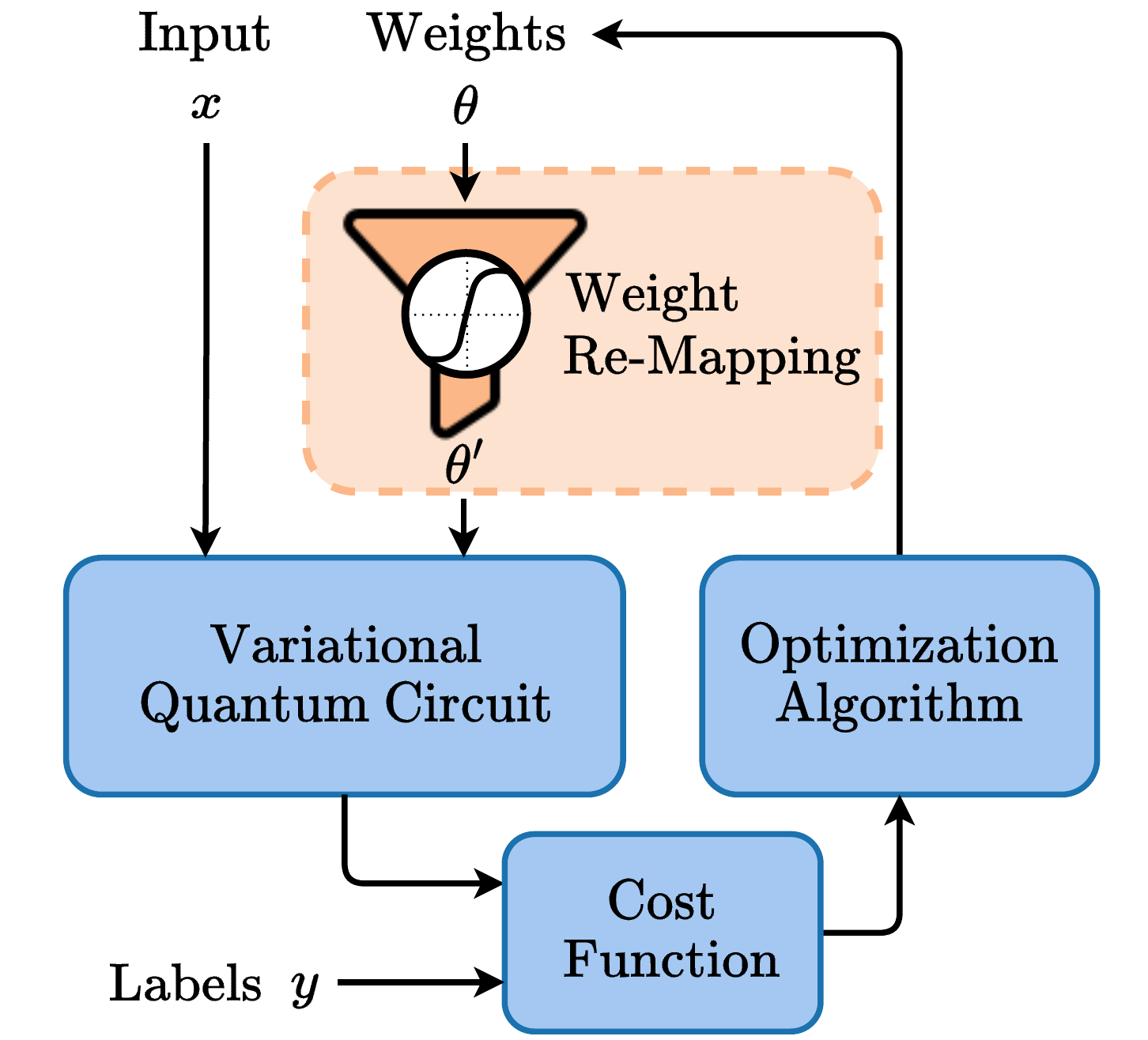}
    \caption{Overview of the variational quantum circuit training process with weight re-mapping by Kölle et al. \cite{koelle2023}}
    \label{fig:structure}
\end{figure}

\noindent Aside from the previously discussed quantum advantage, QML, compared to its classical counterpart, benefits from the need for far fewer parameters in its function approximators \cite{Du_2020}. However, there are still many issues and difficulties that need to be studied and properly addressed when dealing with quantum parameterized circuits, such as the barren plateau phenomena or the fact that adding more layers of gates increases the noise. Another major issue is the fact that the natural domain to which the rotational weights belong is the periodic interval $[0,2\pi]$. The best way to deal with the periodicity of the weights, considering that they are updated with a standard gradient update rule and thus could assume ambiguous values, remains unclear.
Similar to how data re-scaling or normalization techniques have significantly improved performance in classical ML cases \cite{Singh2019}, an approach has been introduced to re-map the parameters of a Variational Quantum Circuit, as illustrated in Figure~\ref{fig:structure} \cite{koelle2023}. More specifically, weight re-mapping functions have been introduced in the architecture of a variational quantum circuit to re-scale the parameters from the real domain to the compact interval $[\pi, \pi]$. 
In this work, we not only elaborate on this approach but also expand upon it by studying a broader range of weight re-mapping functions and testing them on a larger set of datasets. 
Our experimental results demonstrate that using a VQC equipped with a weight re-mapping function invariably guarantees a faster convergence compared to a VQC where no such function is employed. \\

\noindent In this study, we begin with a review of relevant literature and an explanation of the concept of weight re-mapping. This is followed by a detailed exposition of our research methodology and experimental design. Subsequently, we present our findings and engage in a comprehensive discussion of the results. The paper concludes with a summary of the study, an examination of its limitations, and suggestions for future research. All experimental data, along with a PyTorch implementation of the weight re-mapping functions utilized in this study, are accessible via this link \footnote{\url{https://github.com/michaelkoelle/qw-map}}.

\section{\uppercase{Related Work}}
This work is an extension of Kölle et al. \cite{koelle2023} which first introduces weight re-mapping for variational quantum circuits. We expand on this approach providing a wider range of evaluation and adding a new re-mapping function. 
Besides that, there are not many works in the literature focusing on on the embedding of data from the classical domain of $\mathbb{R}^n$ in the quantum one $SU(2^n)$. Mostly they rely on local or global mappings \cite{lloyd_quantum_2020} which add an additional classical computation in the process of VQC evaluation. The classical embedding used is also data-specific and it must thus be adapted to every new dataset. 
The standard practice when it comes to local embeddings is to map each dimension of the dataset of interest to an interval, usually as $[\min(\text{data}), \max(\text{data})]\to [0,1]$. This interval is then used to map each value to a rotation axis to the qubit. The different techniques that could be employed are single-qubit axis \cite{stoudenmire_supervised_2016}, multi-qubit embedding \cite{mitarai_quantum_2018} and random linear maps \cite{wilson_quantum_2019,benedetti_parameterized_2019}.

\section{\uppercase{Weight Re-Mapping}}
\label{sec:weight-constraints}
In this section, we first describe the idea behind the weight mapping technique as introduced in \cite{koelle2023} applied to variational quantum circuits. Then we proceed to illustrate the chosen architecture used for the task of classification and the different datasets used to test the VQC. While this study employs weight re-mapping in the context of classification tasks, it's important to underscore that this technique can be readily adapted for other machine learning paradigms, including unsupervised learning and reinforcement learning. \\

\noindent The standard training procedure of a classical neural network consists in updated the multi-dimensional vector $\vec{\theta} \in \mathbb{R}^n$ containing the weights of the connections between each neuron. Through the back-propagation the gradient of the loss function $\mathcal{L}$ con be computed and the weights are updated according to \ref{update_rule}.

\begin{equation}\label{update_rule}
    \vec{\theta}_i = \vec{\theta}_{i-1} - \alpha \nabla_{\vec{\theta}_{i-1}} \mathcal{L}(\vec{\theta}_{i-1})
\end{equation}

\noindent This represents the classical update step, which implicitly assumes that the space to which the weights belong and in which we follow the gradient direction is $\mathbb{R}^n$. This is true for a classical neural network where a weight can take any value of the real line $\mathbb{R}$. The same doesn't hold for a variational quantum circuit, where a parameter encodes a rotation around a specific axis of the Bloch sphere. 
Although technically every value of the the real line could be assumed by a parameter, we must remember that a rotation of angle $\theta$ has a period of $2\pi$, so that $R(\hat{v}, \theta) = R(\vec{v}, \theta + 2\pi)$. The whole $\mathbb{R}$ domain is thus redundant and letting the parameter take any real number may result in undesired behaviors, since moving according to the gradient of the loss function may lead the parameter in an adjacent interval, meaning that actually we are moving in the opposite direction with respect to the desired one. 
For these reasons it is natural to select a periodic interval $[\phi, \phi + 2 \pi ]$ as the parameter space of a variational quantum circuit. $\phi \in \mathbb{R}$ is a generic phase, and we choose $\phi = -\pi$ so that the chosen interval is centered around 0 and the parameters of a VQC result in $\vec{\theta} \in [-\pi, \pi]^n$.
The weight re-mapping technique is employed to address the described mismatch between classical and quantum weights, so to constrain the classical real weights into the periodic interval. More in general a \textit{mapping function} $\varphi$ behaves by mapping the real line into a compact interval as
    
\begin{equation}
    \varphi : \mathbb{R} \to [a, b]
\end{equation}

\noindent More specifically, according to what has been discussed above, we choose $ a = -b = - \pi$, so that $\varphi : \mathbb{R} \to [-\pi, \pi]$. 
The weight mapping function is introduced in such a way that, before a forward pass is performed, every rotation parameter inside the VQC will first be remapped. We can thus picture every rotation gate of the architecture working as follows

\begin{center}
    \begin{quantikz}[column sep=1cm]
    & \gate{R(\varphi(\theta))} &  \qw
    \end{quantikz}
\end{center}

\noindent Only during the forward pass are thus the weights constraint to the periodic interval. Otherwise they are indeed free to take any value in the real domain. The update process will thus be adapted as 

\begin{equation}
    \vec{\theta}_i = \vec{\theta}_{i-1} - \alpha \nabla_{\vec{\theta}_{i-1}} \mathcal{L}( \varphi( \vec{\theta}_{i-1} ) )
\end{equation}

\noindent which stresses the fact that during the computation of the loss function $\mathcal{L}$, so during the forward pass of the VQC, the weight mapping function $\varphi$ is indeed applied. It is not though during the updated step.\\

\subsection{Re-Mapping Functions}

\begin{figure*}
     \centering
     \begin{subfigure}[t]{0.24\textwidth}
         \centering
         \includegraphics[width=\textwidth]{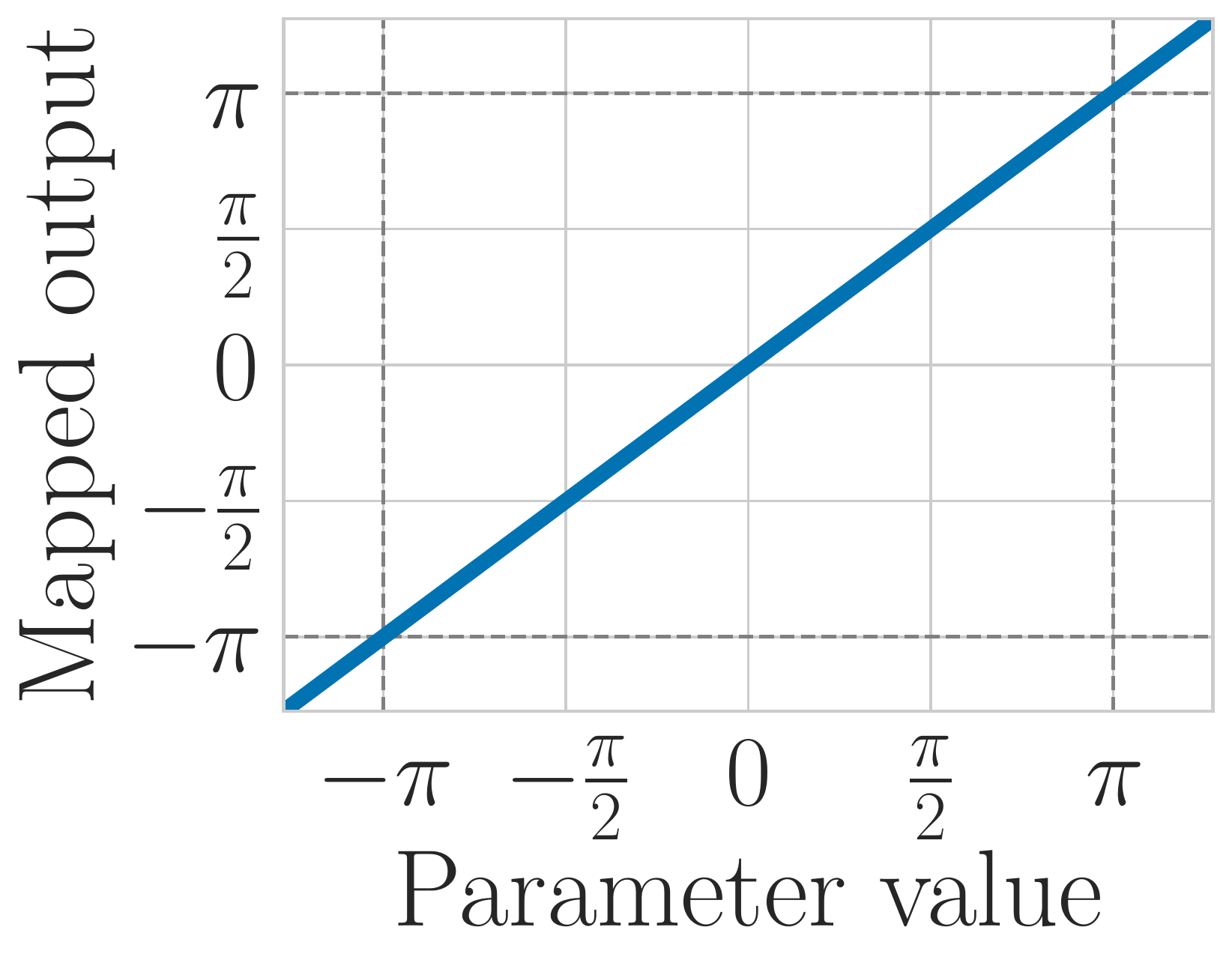}
         \caption{no Re-Mapping}
         \label{fig:func-wo-constraint}
     \end{subfigure}
     \begin{subfigure}[t]{0.24\textwidth}
         \centering
         \includegraphics[width=\textwidth]{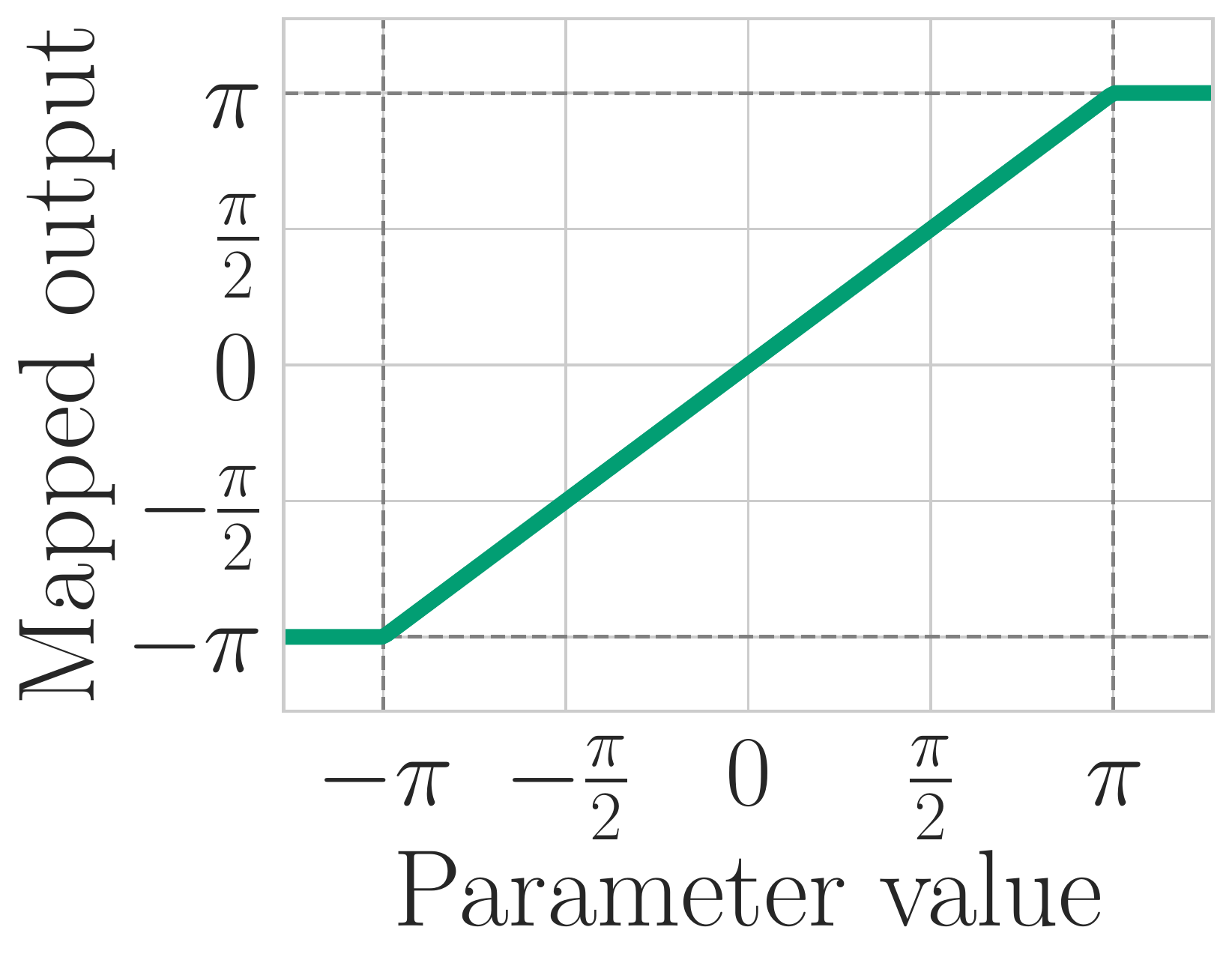}
         \caption{Clamp}
         \label{fig:func-clamp}
     \end{subfigure}
     \begin{subfigure}[t]{0.24\textwidth}
         \centering
         \includegraphics[width=\textwidth]{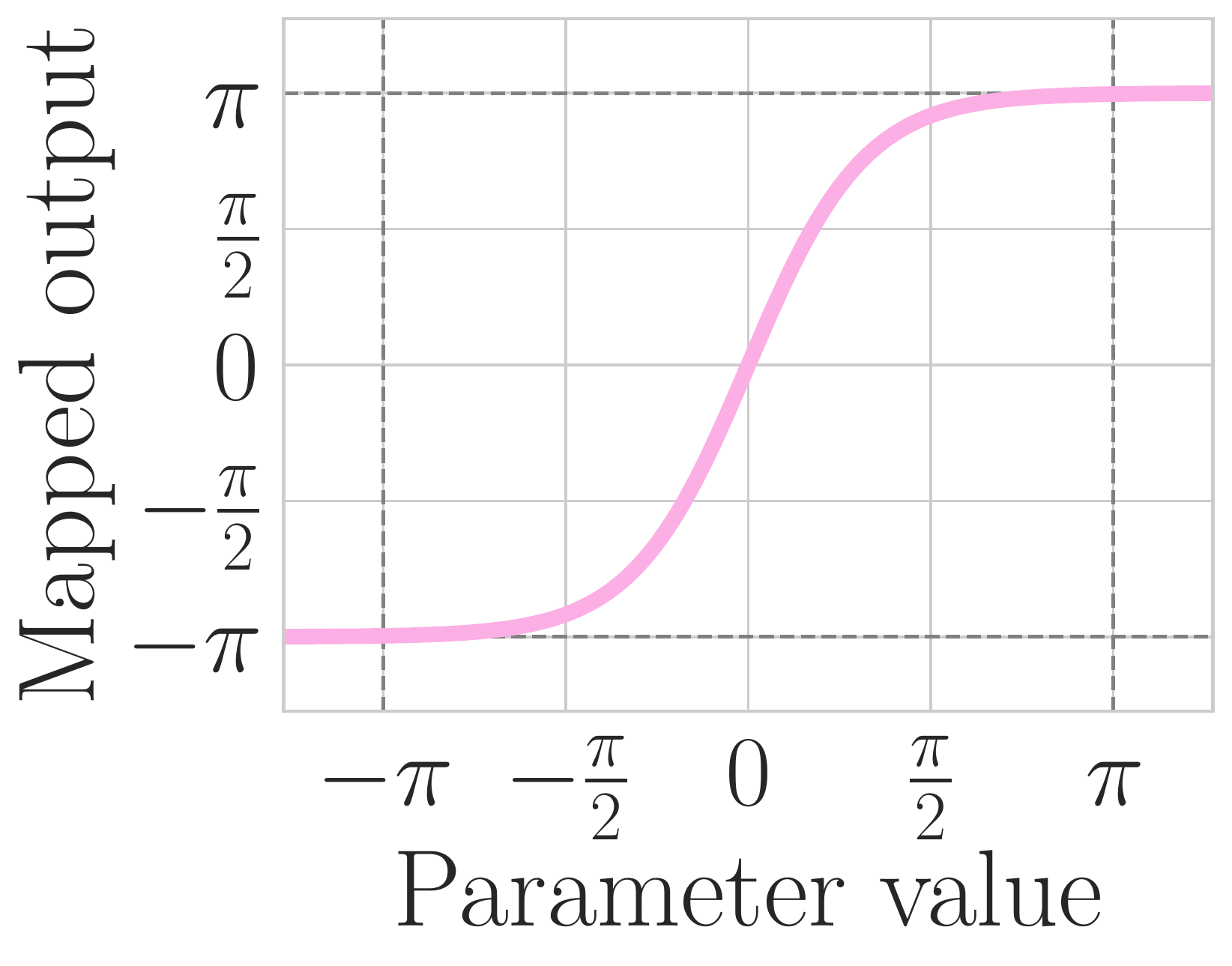}
         \caption{Tanh}
         \label{fig:func-tanh}
     \end{subfigure}
     \begin{subfigure}[t]{0.24\textwidth}
         \centering
         \includegraphics[width=\textwidth]{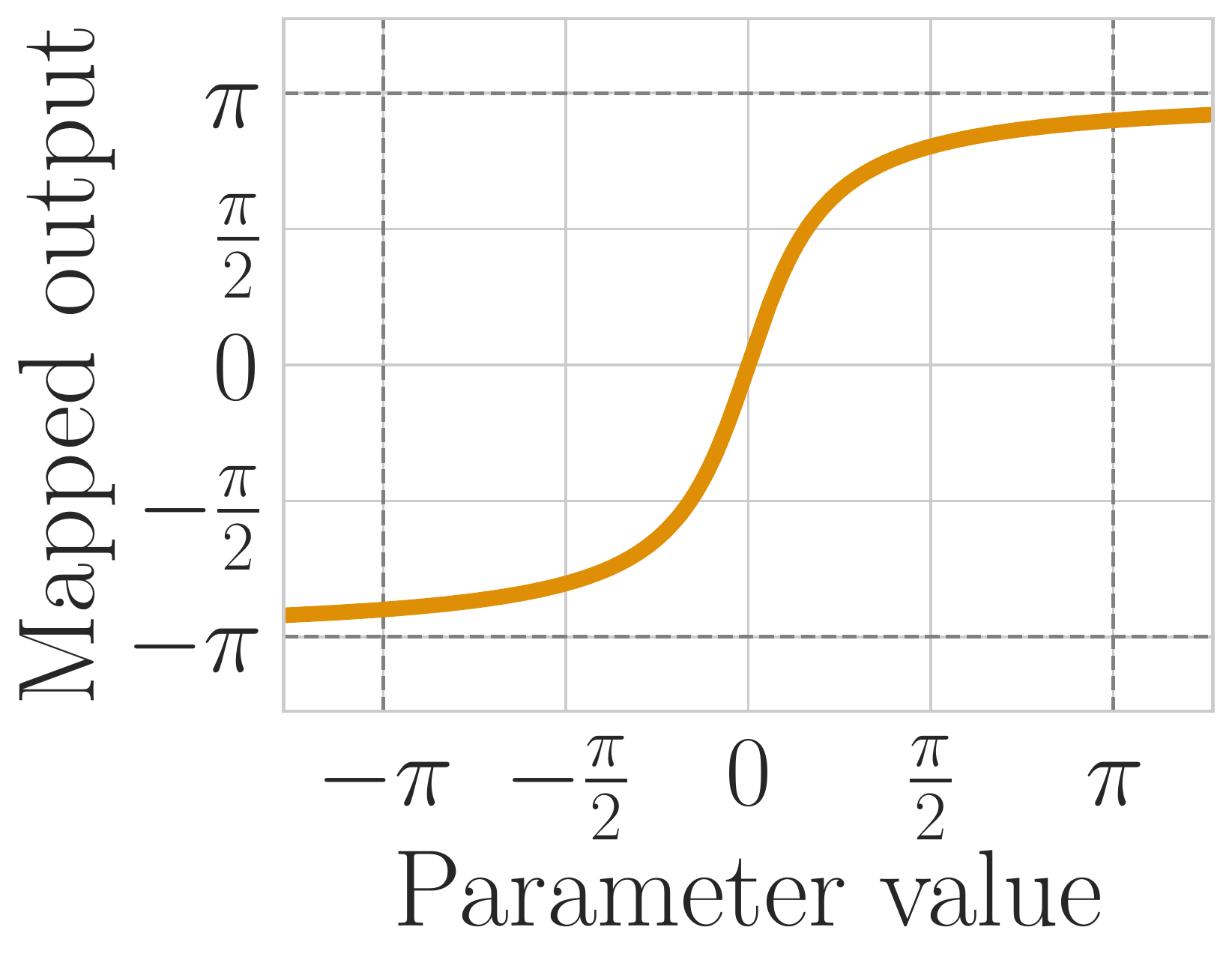}
         \caption{Arctan}
         \label{fig:func-arctan}
     \end{subfigure}\\
     \vspace*{3mm}
     \begin{subfigure}[t]{0.24\textwidth}
         \centering
         \includegraphics[width=\textwidth]{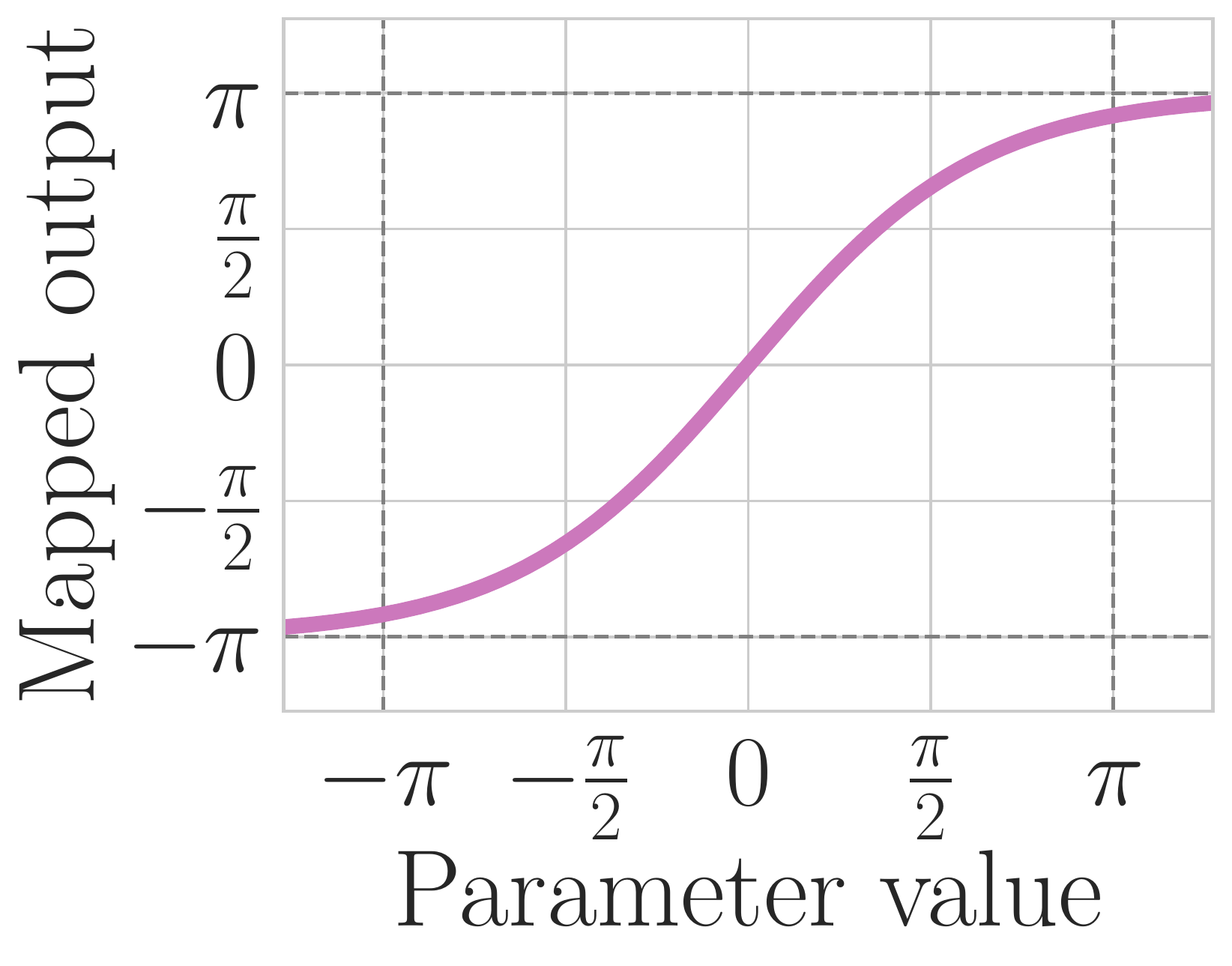}
         \caption{Sigmoid}
         \label{fig:func-sigmoid}
     \end{subfigure}
     \begin{subfigure}[t]{0.24\textwidth}
         \centering
         \includegraphics[width=\textwidth]{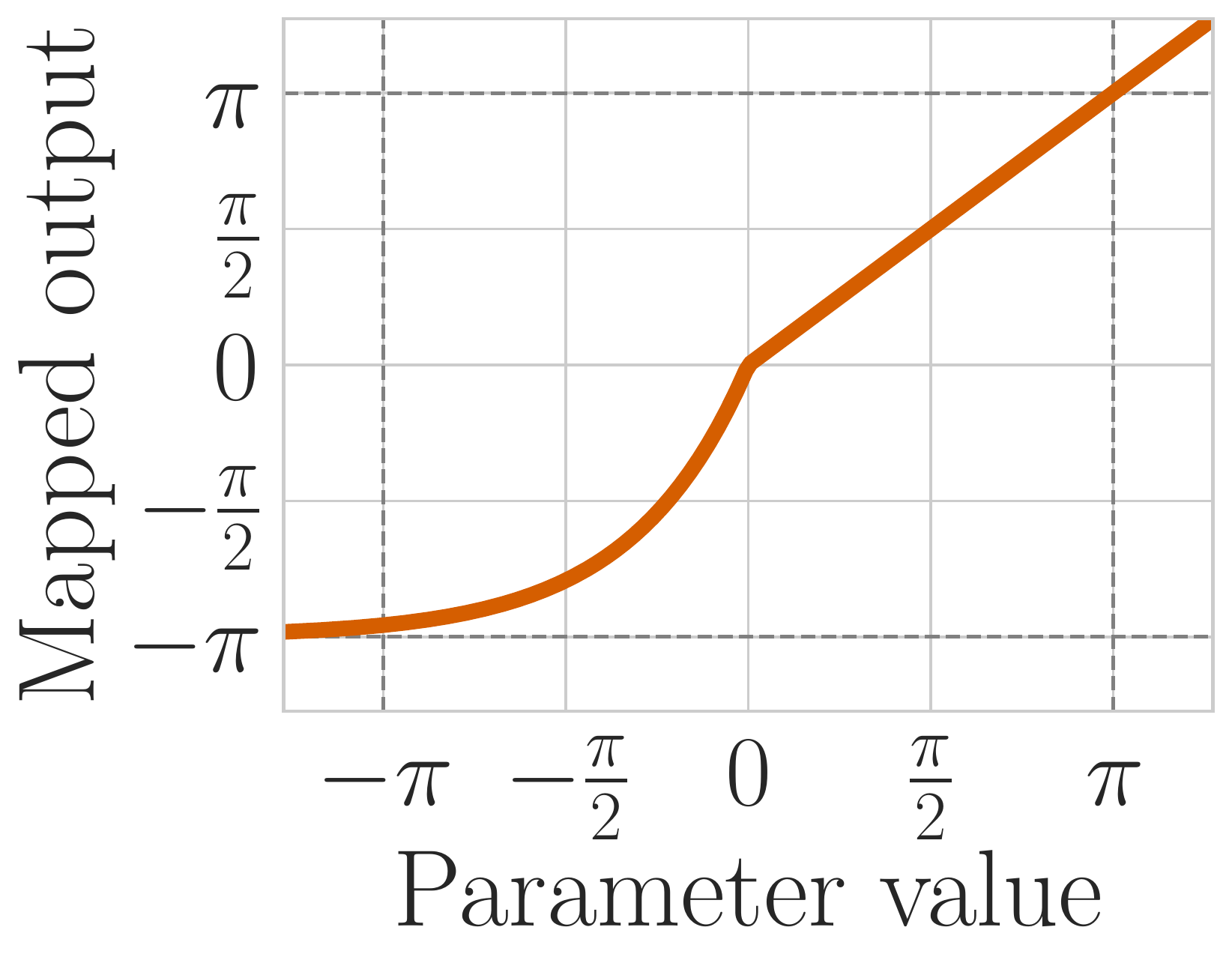}
         \caption{ELU}
         \label{fig:func-elu}
     \end{subfigure}
     \begin{subfigure}[t]{0.24\textwidth}
         \centering
         \includegraphics[width=\textwidth]{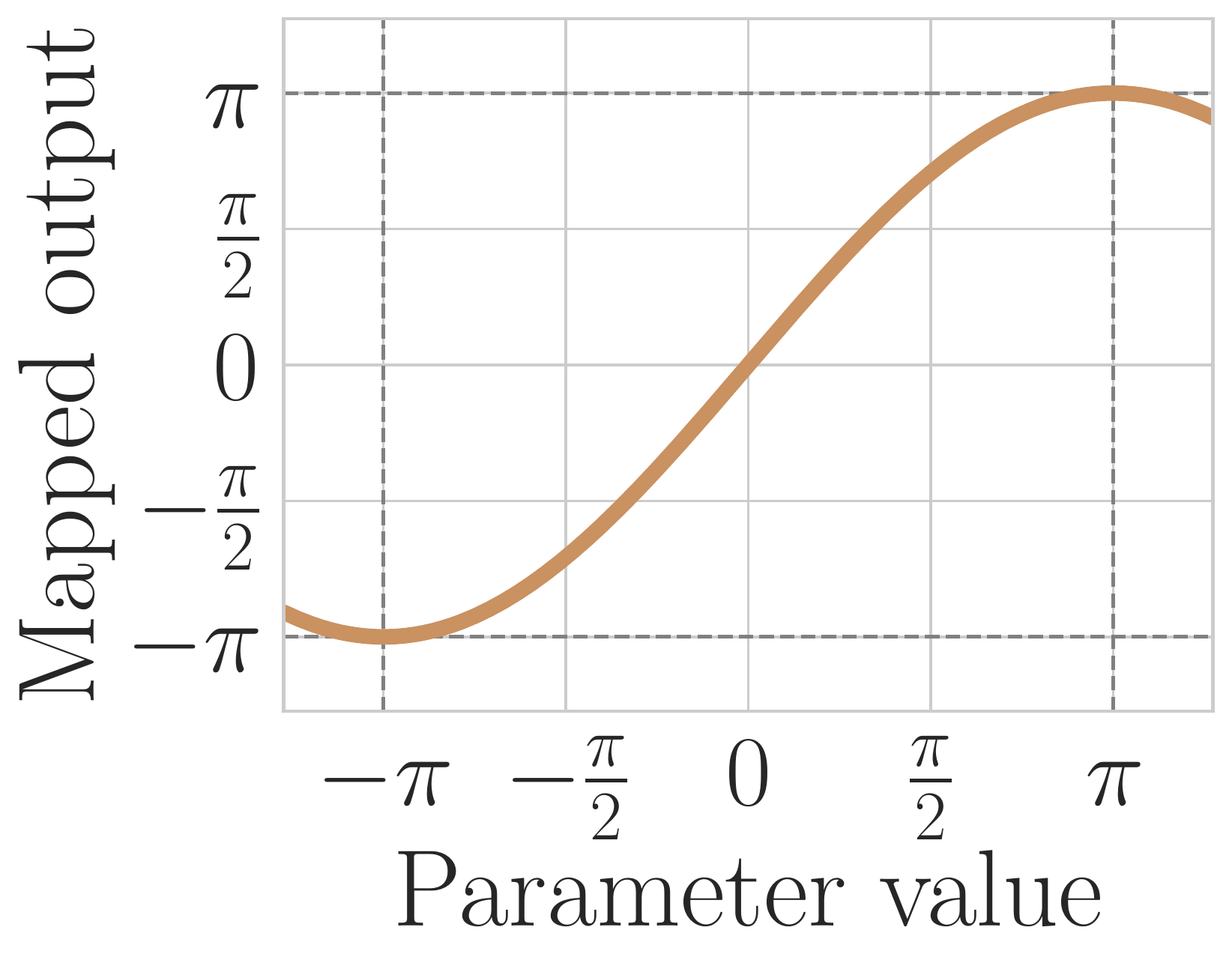}
         \caption{Sinus}
         \label{fig:func-sinus}
     \end{subfigure}
        \caption{Weight re-mapping functions \cite{koelle2023}}
        \label{fig:mapping-functions}
\end{figure*}

\noindent We tested seven mapping functions to see which embedding into the periodic domain leads to better performances. Equations \ref{eq:id},\ref{eq:hard_clipping}, \ref{eq:tanh}, \ref{eq:arctan}, \ref{eq:logistic}, and \ref{eq:elu} were taken from Kölle et al \cite{koelle2023}. Similarly to Kölle et al \cite{koelle2023}, we use the identity function as a baseline, which has the same effect as if no function had been applied, as it can be seen in Figure~\ref{fig:func-wo-constraint}.

\begin{equation} \label{eq:id}
    \varphi_1(\theta) = \theta
\end{equation}

\noindent The first mapping function we introduced is the one in Figure~\ref{fig:func-clamp}. Here all the parameters above $\pi$ or below $-\pi$ are clamped to, respectively, $\pi$ and $-\pi$. The values in between the two extrema are instead left untouched. 

\begin{equation} \label{eq:hard_clipping}
    \varphi_2(\theta) = 
        \begin{cases}
            -\pi & \text{if $\theta < -\pi$}\\
            \pi &\text{if $\theta > \pi$}\\
            \theta &\text{otherwise}
        \end{cases}
\end{equation}

\noindent The next function we selected is the hyperbolic tangent, which is steeper around the $0$ value but smoother around the values $-\pi$ and $\pi$, as shown in Figure \ref{fig:func-tanh}. This means that the values around the bounds of the interval will not be widely differentiated.  

\begin{equation} \label{eq:tanh}
    \varphi_3(\theta) = \pi \tanh(\theta)
\end{equation}

\noindent To conduct further studies on how the steepness of the mapping function affects the embedding of the real weights into the period, we scale the inverse of the tangent function by a factor of 2. The graph is shown in Figure \ref{fig:func-arctan}.

\begin{equation} \label{eq:arctan}
    \varphi_4(\theta) = 2 \arctan(2 \theta)
\end{equation}

\noindent We then tested a function with a steepness in between the two previous one, namely the sigmoid function. This needs obviously to first be rescaled with a factor of $2\pi$ in such a way that the bounds match the $[-\pi, \pi]$ interval. The result can be seen in Figure~\ref{fig:func-sigmoid}.

\begin{equation} \label{eq:logistic}
    \varphi_5(\theta) = \frac{2\pi}{1+e^{-\theta}} -\pi
\end{equation}

\noindent The Exponential Linear Unit (ELU) has been tested as an example of an asymmetric function. To make it converge at least on the lower bound we set $\alpha = \pi$. The result can be seen in Figure \ref{fig:func-elu}.  

\begin{equation} \label{eq:elu}
    \varphi_6(\theta) = 
        \begin{cases}
            \pi(e^\theta-1) &\text{if $\theta < 0$}\\
            \theta &\text{otherwise}
        \end{cases}
\end{equation}

\noindent Lastly we introduce a new re-mapping function based on sinus. The graph is shown in Figure \ref{fig:func-sinus} It has been properly scaled in such a way that its period is $4 \pi$ and its amplitude $2\pi$.

\begin{equation} \label{eq:sin}
    \varphi_7(\theta) = \pi \sin {\frac{\theta}{2}}
\end{equation}

\subsection{Variational Circuit Architecture}
\label{sec:variational-classifier}
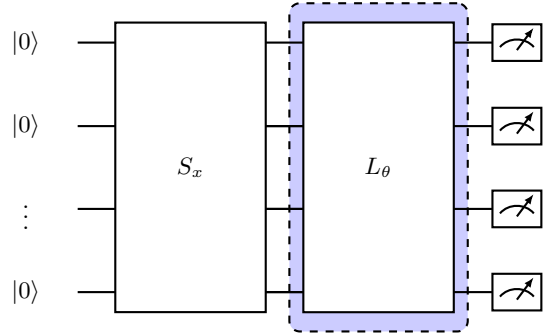
\begin{figure}
    \centering
    \begin{quantikz}
        \ket{0} &  & \gate[wires=4][2cm]{S_x} & \gate[wires=4][2cm]{L_\theta} \gategroup[4,steps=1,style={dashed,
rounded corners,fill=blue!20, inner xsep=2pt},
background]{{\sc }}& \meter{} \\
       \ket{0} & & & \targ{} & \meter{} \\
       \vdots &  & & \qw & \meter{} \\
       \ket{0} & & & \qw & \meter{}
    \end{quantikz}
    
    \caption{Abstract variational quantum circuit used in this work. Dashed blue area indicates repeated layers. \cite{koelle2023}}
    \label{fig:abstract_circuit}
\end{figure}

For the variational classifier used in our study, we have designed a three-component variational quantum circuit, illustrated in Figure \ref{fig:abstract_circuit}. Next, we will discuss this standard design with its three parts individually.

\subsubsection{State preparation}
The first component of our variational quantum circuit handles the state preparation $S_x$, which embeds a feature vector into a quantum circuit. The feature vector is composed of real values from the Euclidean space, which become mapped into the Hilbert space during the embedding process. The most common and general-purpose embedding techniques are the Angle Embedding and Amplitude Embedding. \\

\noindent \textit{Angle Embedding} is a technique used to encode $n$ real-valued entries of the feature vector into the $n$ qubits we use. This encoding starts by initializing the qubits as $\ket{0}$, followed by the embedding of information from the feature vector through the rotation angles, using rotational Pauli-X and Pauli-Y gates. The applied angles thereby represent the feature amplitudes to be encoded, illustrated below, using a single-axis rotational gate.

\begin{center}
    \begin{quantikz}[column sep=1cm]
    & \gate{R_i(x_j)} &  \qw
    \end{quantikz}
\end{center}

\noindent The angle $x_j \in \mathbb{R}$ represents the $j$-th element of feature vector to encode, while the gate $R_i, \text{ with } i \in \{X, Y\}$, represents the rotational gate applied on the $i$-th axis. The application of these rotational gates perform the respective embedding process. A rotation around the $z$-axis is unnecessary here, as the qubits were initialized with $\ket{0}$, a vector along the $z$-axis, which cancels out any effect of such a rotation. All our experiments with Angle Embedding are done using RX-Gates.  \\

\noindent \textit{Amplitude Embedding} consists in embedding the features into the amplitudes of the qubits. Since we can make use of superposition, we can embed $2^n$ features into $n$ qubits. For Example, if we want to embed feature vector $x \in \mathbb{R}^3$ in to a 2 qubit quantum state $\ket{\psi} = \alpha\ket{00} + \beta\ket{01} + \gamma\ket{10} + \delta\ket{11}$ such that $\lvert \alpha\rvert^2 + \lvert \beta\rvert^2 + \lvert \gamma\rvert^2 + \lvert \delta\rvert^2 = 1$, we need to first pad our feature vector so that it matches $2^n$ features where $n$ is the number of qubits used. Next, we need to normalize the padded feature vector $y$ such that $\sum_{k=0}^{2^n - 1} \frac{y_k}{\lvert\lvert y\rvert\rvert} = 1$. Lastly, we use a technique like the state preperation by Mottonen et al. \cite{mottonen2004transformation} to embed the padded and normalized feature vector into the amplitudes of the qubits state.

\subsubsection{Variational Layers}

After conducting the quantum state preparation, the circuit implements iteratively layers of gates. Each layer, $L_\theta$, consists of three single qubit rotations, as well as CNOT gates, used as entanglers. The number of iteratively implemented layers is determined by the layer-count $L$. In Figure \ref{fig:abstract_circuit} the blue dashed area represents a single layer of gates which is then repeated $L$ times. The implementation was inspired by the circuit-centric classifier design \cite{Schuld_2020}, exemplary shown as three qubit implementation below. 

\begin{center}
\begin{adjustbox}{width=0.7\linewidth}
\begin{quantikz}
     \qw& \gate{RZ(\theta_0^0)} & \gate{RY(\theta_0^1)}& \gate{RZ(\theta_0^2)} & \ctrl{1} & \qw      & \targ{} & \qw   \\
     \qw&\gate{RZ(\theta_1^0)} & \gate{RY(\theta_1^1)}& \gate{RZ(\theta_1^2)} & \targ{}  & \ctrl{1} & \qw    & \qw    \\
     \qw&\gate{RZ(\theta_2^0)} & \gate{RY(\theta_2^1)}& \gate{RZ(\theta_2^2)} & \qw      & \targ{}  & \ctrl{-2}& \qw 
\end{quantikz}
\end{adjustbox}
\end{center}

\noindent The trainable parameters $\theta_i^j$, get applied on the $i$-th qubit and the $j$-th rotational gate, whereby $j \in \{0,1,2\}$. For simplicity, the shown graphic does omit the index of the respective layer. For the CNOT gates, responsible for the entanglement, the target bit is calculated by $(i + l)\ \text{mod}\ n$, which reveals the dependency on the respective layer-number $l$. This way of construction ensures for layer $l=1$ a circular entanglement, as the control and target qubits are direct neighbors, exceptional of the last CNOT-gate, which constructs the circular character of entanglement. For the second layer, $l=2$, control and target qubits are partly separated by another qubit. The index of the respective target qubit of the first qubit ($i=0$) is here the third qubit ($i=2$).

\subsubsection{Measurement}
Looking at our circuit architecture, the last part of the variational circuit consists of a measurement layer. Here, the expectation values of the first $k$ qubits, where $k$ depends on the classes to determine, is measured in their computational basis ($z$). Afterwards, we add a bias to each of the measured expectation values, as those biases are also part of the parameters and updated accordingly. Ending the experiment, the softmax of the measurement is computed to normalize the respective probabilities. 

\subsection{Data Re-Uploading Architecture}
\begin{figure}
    \centering
    \begin{quantikz}
        \ket{0} &  & \gate[wires=4][2cm]{S_x} \gategroup[4,steps=2,style={dashed,
rounded corners,fill=blue!20, inner xsep=2pt},
background]{{\sc }} & \gate[wires=4][2cm]{L_\theta} & \meter{} \\
       \ket{0} & & & \targ{} & \meter{} \\
       \vdots &  & & \qw & \meter{} \\
       \ket{0} & & & \qw & \meter{}
    \end{quantikz}
    
    \caption{Variational quantum circuit with data re-uploading used in this work. Dashed blue area indicates repeated encoding and layers.}
    \label{fig:abstract_data_reup_circuit}
\end{figure}
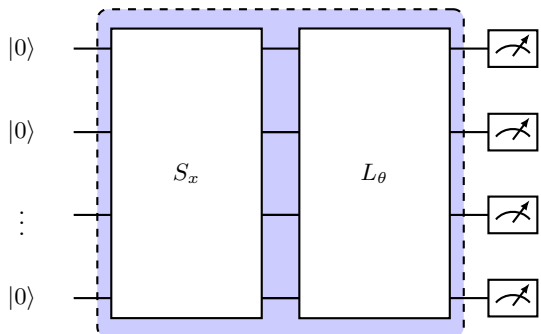
Originally introduced in Pérez-Salinas et al. \cite{P_rez_Salinas_2020}, data re-uploading gets its name from the fact that the data is re-uploaded into the quantum circuit multiple times, at each layer of the circuit. This allows for a more expressive quantum circuit, as it can generate a richer set of transformations on the input data. The data re-uploading strategy is particularly useful for variational quantum classifiers because it allows the quantum circuit to learn complex, non-linear decision boundaries. This process effectively increases the expressivity of the quantum circuit, enabling it to capture more complex patterns in the data.
In the context of our work, we utilize data re-uploading to enhance the performance of our variational quantum classifier. By re-uploading the data at each layer of the circuit, we are able to construct a more expressive quantum model that can better capture the intricacies of our dataset. This is particularly important for our investigation into the performance of the weight re-mapping method on more expressive circuits, as it allows us to explore the full potential of this technique.
In terms of circuit construction, a circuit with data re-uploading is built from alternating embedding and variational layers (Figure \ref{fig:abstract_data_reup_circuit}). The embedding layers, denoted as $S_x$, are responsible for mapping the classical data onto the quantum state space, while the variational layers, denoted as $L_\theta$, apply a series of parameterized quantum gates that transform the quantum state. By alternating between these two types of layers, we are able to repeatedly re-upload and transform the data, thereby increasing the expressivity of the circuit.The structure of a data re-uploading circuit can be represented as follows:

\begin{equation}
U(\theta, x) = L_{\theta_n} S_{x_n} \ldots L_{\theta_1} S_{x_1}
\end{equation}

\noindent where $U(\theta, x)$ is the total unitary operation representing the data re-uploading circuit, $\theta$ and $x$ are the sets of parameters and input data respectively, and $n$ is the number of layers in the circuit.

\section{\uppercase{Experimental Setup}}
In the Experimental Setup section, we detail the methodology employed to evaluate the performance of the variational classifier, as described in section \ref{sec:variational-classifier}. This includes the datasets used for testing, the baselines against which the classifier was compared, the metrics used to assess performance, and the hyperparameters and training procedures applied.

\subsection{Datasets}
In this section, we present the datasets that were used to assess the variational classifier from section \ref{sec:variational-classifier}. We chose two popular datasets that pose a simple supervised learning classification task.

\subsubsection{Abalone Dataset}
\label{sec:abalone}
The Abalone dataset is a well-known dataset in the field of machine learning and data mining. It was created from physical measurements of abalones, a type of marine snail, with the goal of predicting the age of the abalone from these measurements \cite{abalone_dataset}. The dataset includes three classes, each representing a different sex of the abalone: Male, Female, and Infant. Each feature vector in the dataset contains eight attributes: length, diameter, height, whole weight, shucked weight, viscera weight, shell weight, and rings. The Abalone dataset is not balanced. The number of instances for each class are as follows: Male: 1528, Female: 1307, and Infant: 1342.

\subsubsection{Banknote Authentication Dataset}
\label{sec:banknote}
The Banknote Authentication dataset is a popular dataset for binary classification tasks. It was created from images of genuine and forged banknote-like specimens, using an industrial camera usually used for print inspection \cite{Dua:2019}. The images were then processed using a wavelet transform tool to extract features from them. The dataset includes two classes, representing genuine and forged banknotes. Each feature vector in the dataset contains four attributes: variance of Wavelet Transformed image, skewness of Wavelet Transformed image, curtosis of Wavelet Transformed image, and entropy of image. These features capture the essential characteristics of the banknote images, making them suitable for the task of banknote authentication. The Banknote Authentication dataset roughly balanced, with 762 instances for the genuine class and 610 instances for the forged class.

\subsubsection{Glass Identification Dataset}
\label{sec:glass}
The Glass Identification dataset is a multi-class classification dataset that was created from the analysis of different types of glass. The dataset was collected for the purpose of forensic science to identify the type of glass found at crime scenes \cite{Dua:2019}. The dataset includes seven classes, each representing a different type of glass: building windows float processed, building windows non-float processed, vehicle windows float processed, containers, tableware, headlamps, and an undefined category. Each feature vector in the dataset contains nine attributes: refractive index, Sodium (Na), Magnesium (Mg), Aluminum (Al), Silicon (Si), Potassium (K), Calcium (Ca), Barium (Ba), and Iron (Fe). These features represent the chemical composition of the glass, which varies depending on the type of glass. The Glass Identification dataset is not balanced. The number of instances for each class are as follows: building windows float processed: 70, building windows non-float processed: 76, vehicle windows float processed: 17, containers: 13, tableware: 9, headlamps: 29, and undefined: 6.

\subsubsection{Heart Disease Dataset}
\label{sec:heart}
The Heart Disease dataset is a widely used dataset in the field of medical informatics and machine learning. It was created from several different studies and has been used in numerous research papers to develop and test algorithms for binary classification \cite{heart}. The dataset includes two classes, representing the presence and absence of heart disease. Each feature vector in the dataset contains 13 attributes: age, sex, chest pain type, resting blood pressure, serum cholesterol, fasting blood sugar, resting electrocardiographic results, maximum heart rate achieved, exercise induced angina, ST depression induced by exercise relative to rest, the slope of the peak exercise ST segment, number of major vessels colored by fluoroscopy, and thalassemia. The Heart Disease dataset is balanced, with 165 instances for the class representing the presence of heart disease and 138 instances for the class representing the absence of heart disease.

\subsubsection{Iris Dataset}
\label{sec:iris}
The Iris dataset, first introduced in an article by Fisher in 1936 \cite{fisher1936use}, has since become a standard benchmark in the field of classification, and continues to be widely used in literature. For this study, we have used an updated version of the dataset, which corrects two minor discrepancies found in the original publication. This dataset is composed of three distinct classes, each representing a type of iris plant: Iris Setosa, Iris Versicolour, and Iris Virginica. Each class contains 50 instances, where every instance is characterized by a four-dimensional feature vector including sepal length, sepal width, petal length, and petal width (all measurements are in centimeters). While the first two classes (Setosa and Versicolour) can be linearly separated, the latter two (Versicolour and Virginica) are not linearly separable, posing a greater challenge for classification. We also use a version with the first two classes, resulting in a easier classification task.

\subsubsection{Pima Indians Diabetes Dataset}
\label{sec:pima}
The Pima Indians Diabetes dataset is a widely used dataset in the field of medical informatics and machine learning. It was created by the National Institute of Diabetes and Digestive and Kidney Diseases and aims to predict whether or not a patient has diabetes based on certain diagnostic measurements \cite{diabetes}. The dataset includes two classes, representing the presence and absence of diabetes. Each feature vector in the dataset contains eight attributes: number of times pregnant, plasma glucose concentration, diastolic blood pressure, triceps skin fold thickness, 2-hour serum insulin, body mass index, diabetes pedigree function, and age. The Pima Indians Diabetes dataset is not balanced, with 500 instances for the class representing the absence of diabetes and 268 instances for the class representing the presence of diabetes.

\subsubsection{Seeds Dataset}
\label{sec:seeds}
The Seeds dataset is a commonly used dataset in the field of pattern recognition. It was created from measurements of geometrical properties of kernels belonging to three different varieties of wheat: Kama, Rosa, and Canadian \cite{seeds}. The dataset includes three classes, each representing a different variety of wheat. Each feature vector in the dataset contains seven attributes: area, perimeter, compactness, length of kernel, width of kernel, asymmetry coefficient, and length of kernel groove. The Seeds dataset is balanced, with 70 instances for each class, resulting in a total of 210 instances. 

\subsubsection{Wine Dataset}
\label{sec:wine}
The Wine dataset was produced from a chemical analysis of wines grown in the same region of Italy, which are derived from three different grape varieties \cite{parvus}. This analysis was conducted in July 1991, originally examining 30 constituents. Unfortunately, the comprehensive details of these constituents have been lost over time, and the surviving dataset now comprises 13 chemical properties: Alcohol content, Malic acid, Ash, Alkalinity of ash, Magnesium, Total phenols, Flavanoids, Nonflavanoid phenols, Proanthocyanins, Color intensity, Hue, OD280/OD315 of diluted wines, and Proline. This dataset is not evenly distributed across the three classes, which represent the different grape varieties. The current composition is as follows: class 1 contains 59 instances, class 2 includes 71 instances, and class 3 encapsulates 48 instances.

\subsection{Baselines}
\label{sec:baselines}

In our experiments, we compared our models against the following baselines: a variational quantum classifier (VQC) without re-mapping and, in the last experiment, a classical neural network (Classical NN).

\subsubsection{Variational Quantum Classifier without Re-mapping} The first baseline is a VQC without any re-mapping. This model uses a circuit with 6 layers and a varying embedding method. The embedding methods used include amplitude embedding, angle embedding, and amplitude embedding with data re-uploading. This baseline allows us to evaluate the impact of the re-mapping technique on the performance of the VQC.

\subsubsection{Classical Neural Network} The second baseline is a classical NN, which we only used in the last experiment with the 2-class Iris Dataset. The architecture of this NN consists of an input layer with 4 nodes, a hidden layer with 6 nodes followed by an Exponential Linear Unit (ELU) activation function, and an output layer with a single node followed by a Sigmoid activation function. The choice of 6 nodes in the hidden layer was made to ensure a similar number of parameters to the VQC approach, allowing for a more direct comparison.

\subsection{Metrics}
\label{metrics}
In our experiments, we used two primary metrics besides the loss to evaluate the performance of our models: accuracy and the point of convergence.

\subsubsection{Accuracy} The accuracy is a common metric used in classification tasks. It is defined as the proportion of correct predictions made by the model out of all predictions. In mathematical terms, if $y_i$ is the true label and $\hat{y}_i$ is the predicted label, the accuracy is given by:

\begin{equation}
\text{Accuracy} = \frac{1}{N}\sum_{i=1}^{N} \delta(y_i, \hat{y}_i)
\end{equation}

\noindent where $N$ is the total number of samples, and $\delta(y_i, \hat{y}_i)$ is the Kronecker delta, which is 1 if $y_i = \hat{y}_i$ and 0 otherwise.\\

\noindent We distinguish between training, validation, and test accuracy. Training accuracy is calculated on the same data that was used for training the model. It gives an indication of how well the model has learned the training data, but it does not necessarily reflect how well the model will perform on unseen data. Validation accuracy is calculated on a separate set of data not used in training. It is used to tune hyperparameters and to get an estimate of the model's performance on unseen data while training. Test accuracy is calculated on another separate set of data after the model has been fully trained. It gives an unbiased estimate of the model's performance on unseen data.

\subsubsection{Point of Convergence} The point of convergence (POC) is a metric that indicates when the model's learning process has stabilized. It is defined as the minimum step $t$ at which the difference in validation loss falls below a threshold. The threshold is defined as the product of a factor $k$ and the standard deviation of the validation loss. Mathematically, this is represented as:

\begin{align*}
\text{POC} &= \min_{t} \{t : \Delta \mathcal{L}_{\text{valid}}(\theta_t) < k \cdot \sigma(\mathcal{L}_{\text{valid}}(\theta))\}
\end{align*}

\noindent In our experiments, we chose $k=1$. This choice means that the POC is the point at which the change in validation loss is less than one standard deviation of the validation loss. This is a valid way to determine the point of convergence because it ensures that the model's learning process has stabilized to within the normal variation of the validation loss. It provides a balance between waiting for the model to fully converge and stopping early to prevent overfitting.

\subsection{Hyperparameters and Training}
\label{sec:training}
In all of our experiments, we used a consistent set of hyperparameters to ensure a fair comparison between different approaches. The learning rate was set to $0.01$ and the batch size was set to $5$ for all experiments. For the variational quantum classifier approaches, we used a circuit with 6 layers. The embedding method varied between experiments, with amplitude embedding, angle embedding, and amplitude embedding with data re-uploading being used.  For the classical neural network approach, we used a hidden layer size of 6. This was chosen to ensure a similar number of parameters to the VQC approach, allowing for a more direct comparison. All experiments used the cross entropy loss function and the stochastic gradient descent (SGD) optimizer. Each experiment was repeated 10 times, with seeds ranging from 0 to 9, to ensure the robustness of our results. 
\section{\uppercase{Experiments}}

In this section we present the results of the experiments that we have carried out to benchmark the impact of the weight re-mapping functions. In principle our approach can be applied to any machine learning task relying on a Variational Quantum Circuit, such as Quantum Reinforcement Learning, The Quantum Approximate Optimization Algorithm, and so on. Here we chose supervised classification as our case study: a VQC has been used to classify 8 datasets and the re-mapping functions have been applied to see how the convergence of the classifier is improved. \\

\noindent We trained the classifier on the following datasets: Banknote Authentication, Glass Identification, Heart Desease, Pima Indians Diabetes, Iris, Seeds, Wine and Abalone dataset. The details of the training can be found in Section \ref{sec:training}.
We first show and comment more in detail in Section~\ref{sec:convergence-results} the results relative to the different dataset to see how the functions affect the convergence speed. In doing so, we separate the two different encodings: amplitude and angle embedding. We only comment the top three performing functions. 
Then we show in Section~\ref{sec:average-performance} how each weight re-mapping function affects the convergence performance on average, considering all the datasets at once.
We also studied the test accuracy and we perform an ANOVA test to check if and how the final performance, measured once the training is completed, is affected by the different functions. The results are shown in Section~\ref{sec:test-accuracy}.
In conclusion, as a small comparison with the classical counterpart, a classical neural network has been evaluated and compared with a variational quantum circuit in Section~\ref{sec:classical-comparison}. 

\subsection{Convergence results}\label{sec:convergence-results}

\begin{table}[htb!]
    \begin{adjustbox}{width=0.7\linewidth, center}
\begin{tabular}{lcccccc}
\toprule
approach & VQC-arctan & VQC-clamp & VQC-elu & VQC-sigmoid & VQC-sin & VQC-tanh \\
dataset &  &  &  &  &  &  \\
\midrule
abalone & $\phantom{-}0.016$ & $\phantom{-}0.000$ & $\phantom{-}0.013$ & $-0.003$ & \pmb{$-0.003$} & $\phantom{-}0.016$ \\
banknote & $\phantom{-}0.060$ & $\phantom{-}0.000$ & $\phantom{-}0.008$ & $\phantom{-}0.002$ & $\phantom{-}0.002$ & $\phantom{-}0.019$ \\
glass & $-0.241$ & $\phantom{-}0.000$ & $-0.085$ & $\phantom{-}0.068$ & $\phantom{-}0.068$ & \pmb{$-0.284$} \\
heart\_deasease & \pmb{$-0.123$} & $\phantom{-}0.000$ & $-0.019$ & $-0.086$ & $-0.085$ & $-0.117$ \\
indian\_diabetes & \pmb{$-0.017$} & $\phantom{-}0.000$ & $-0.003$ & $\phantom{-}0.002$ & $\phantom{-}0.000$ & $\phantom{-}0.002$ \\
iris & \pmb{$-0.189$} & $\phantom{-}0.000$ & $-0.009$ & $-0.121$ & $-0.121$ & $-0.141$ \\
seeds & $-0.164$ & $\phantom{-}0.000$ & $-0.235$ & $-0.129$ & $-0.137$ & \pmb{$-0.389$} \\
wine & \pmb{$-0.443$} & $\phantom{-}0.000$ & $-0.070$ & $-0.043$ & $-0.043$ & $-0.419$ \\
\midrule
$\varnothing$ & $-0.138$ & $\phantom{-}0.000$ & $-0.050$ & $-0.039$ & $-0.040$ & \pmb{$-0.164$} \\
\bottomrule
\end{tabular}
    \end{adjustbox}
    \vspace*{3mm}
    \caption{Difference in validation accuracy compared to VQC at point of convergence using Amplitude Embedding}
    \label{tab:amp-convergence}
\end{table}
\begin{table}[htb!]
    \begin{adjustbox}{width=0.7\linewidth, center}
\begin{tabular}{lcccccc}
\toprule
approach & VQC-arctan & VQC-clamp & VQC-elu & VQC-sigmoid & VQC-sin & VQC-tanh \\
dataset &  &  &  &  &  &  \\
\midrule
abalone & \pmb{$-0.165$} & $\phantom{-}0.000$ & $\phantom{-}0.000$ & $\phantom{-}0.000$ & $\phantom{-}0.000$ & $-0.133$ \\
banknote & $-0.228$ & $\phantom{-}0.000$ & $-0.201$ & $-0.235$ & \pmb{$-0.236$} & $-0.227$ \\
glass & \pmb{$-0.247$} & $\phantom{-}0.000$ & $-0.079$ & $-0.151$ & $-0.180$ & $-0.125$ \\
heart\_deasease & \pmb{$-0.166$} & $\phantom{-}0.000$ & $-0.105$ & $-0.126$ & $-0.124$ & $-0.081$ \\
indian\_diabetes & \pmb{$-0.035$} & $\phantom{-}0.000$ & $-0.005$ & $-0.035$ & $-0.035$ & $-0.015$ \\
iris & $-0.222$ & $\phantom{-}0.000$ & $-0.195$ & $-0.124$ & $-0.131$ & \pmb{$-0.282$} \\
seeds & \pmb{$-0.684$} & $\phantom{-}0.000$ & $-0.543$ & $-0.272$ & $-0.277$ & $-0.451$ \\
wine & \pmb{$-0.133$} & $\phantom{-}0.000$ & $\phantom{-}0.075$ & $\phantom{-}0.024$ & $\phantom{-}0.024$ & $-0.022$ \\
\midrule
$\varnothing$ & \pmb{$-0.235$} & $\phantom{-}0.000$ & $-0.132$ & $-0.115$ & $-0.120$ & $-0.167$ \\
\bottomrule
\end{tabular}
    \end{adjustbox}
    \vspace*{3mm}
    \caption{Difference in validation accuracy compared to VQC at point of convergence using Angle Embedding}
    \label{tab:ang-convergence}
\end{table}

\begin{figure}[htb!]
\centering
\subfloat[][Iris Dataset]{\includegraphics[width=0.5\textwidth]{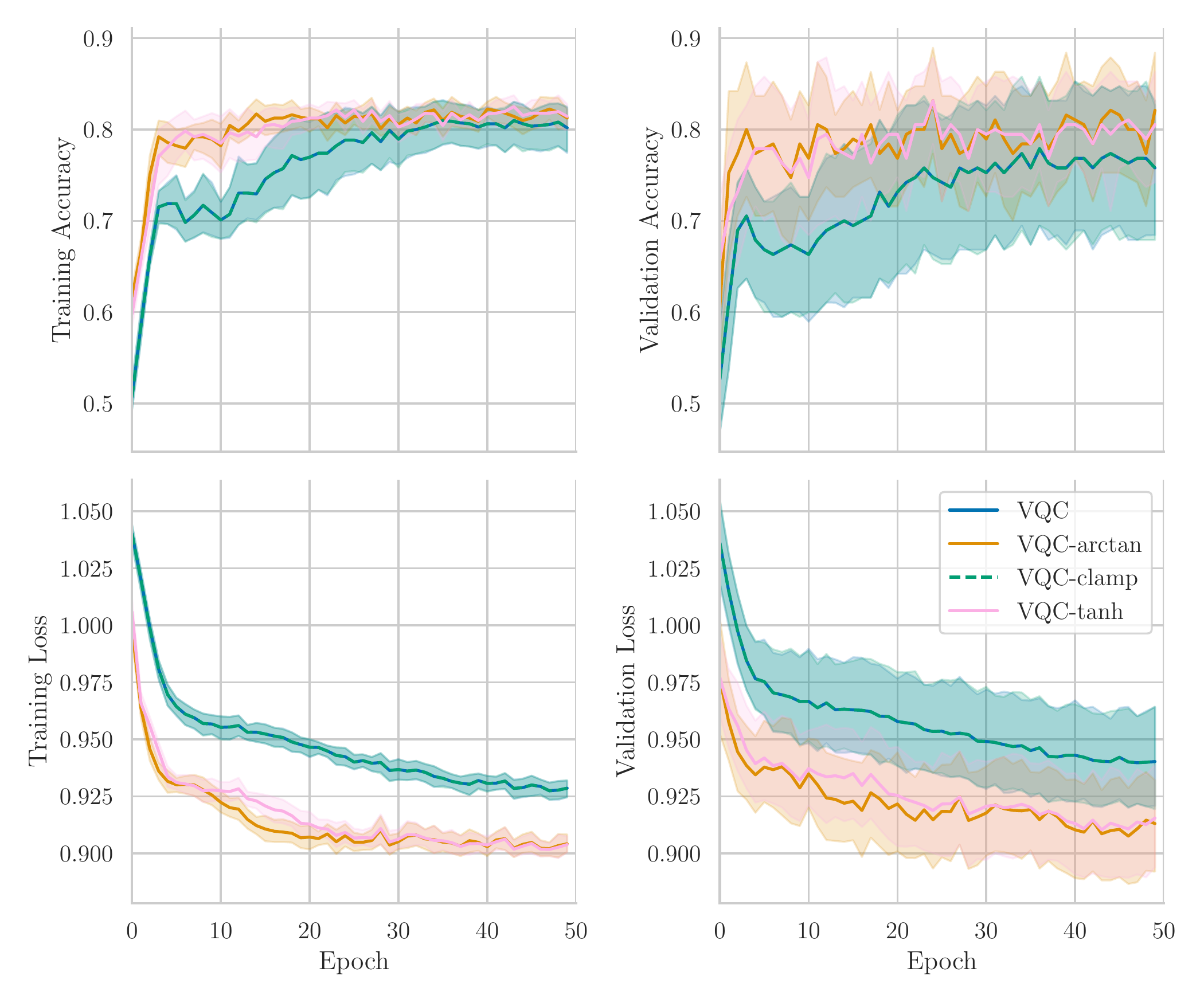}\label{fig:amp-iris}}
\subfloat[][Wine Dataset]{\includegraphics[width=0.5\textwidth]{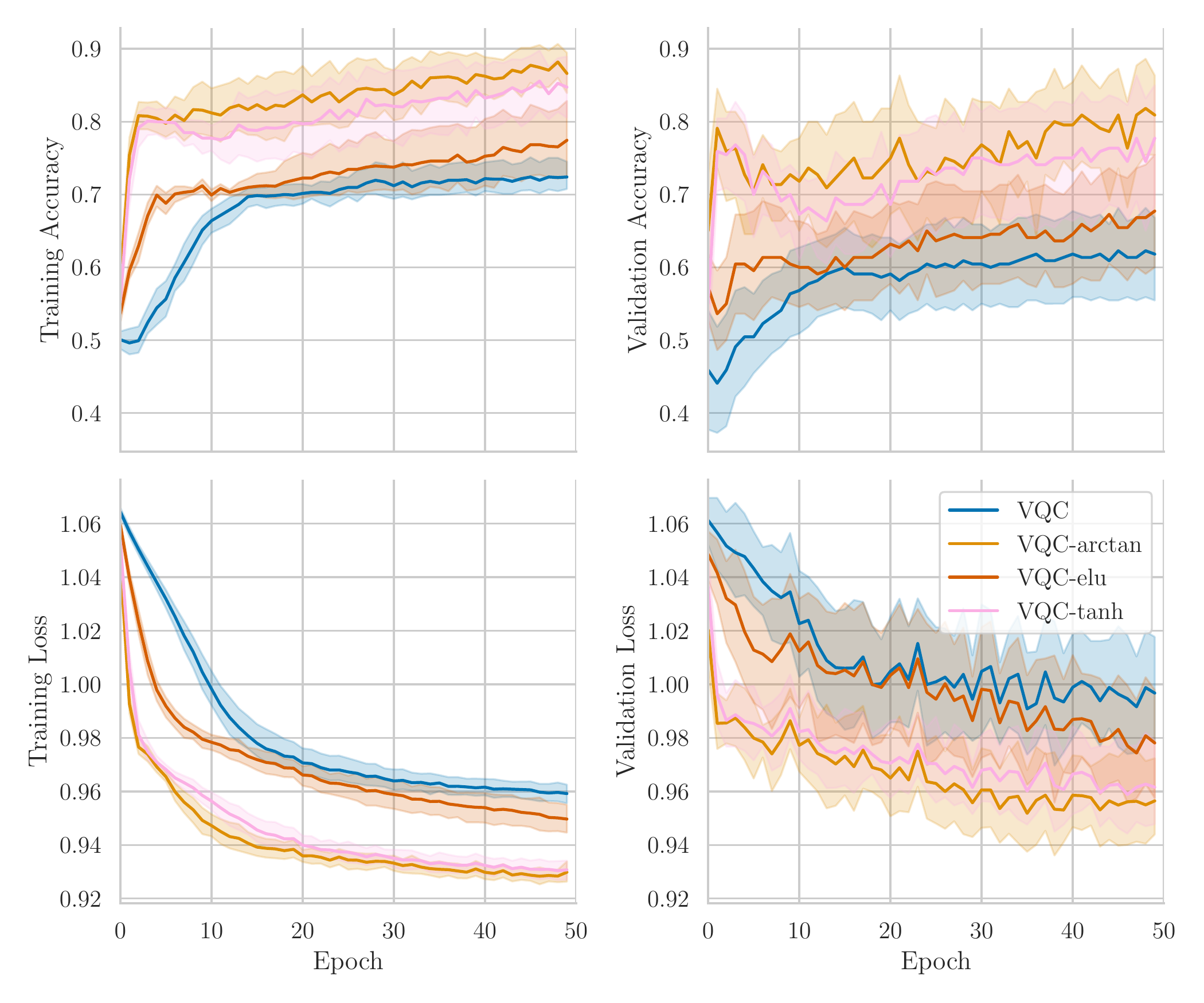}\label{fig:amp-wine}}\\
\subfloat[][Heart Decease Dataset]{\includegraphics[width=0.5\textwidth]{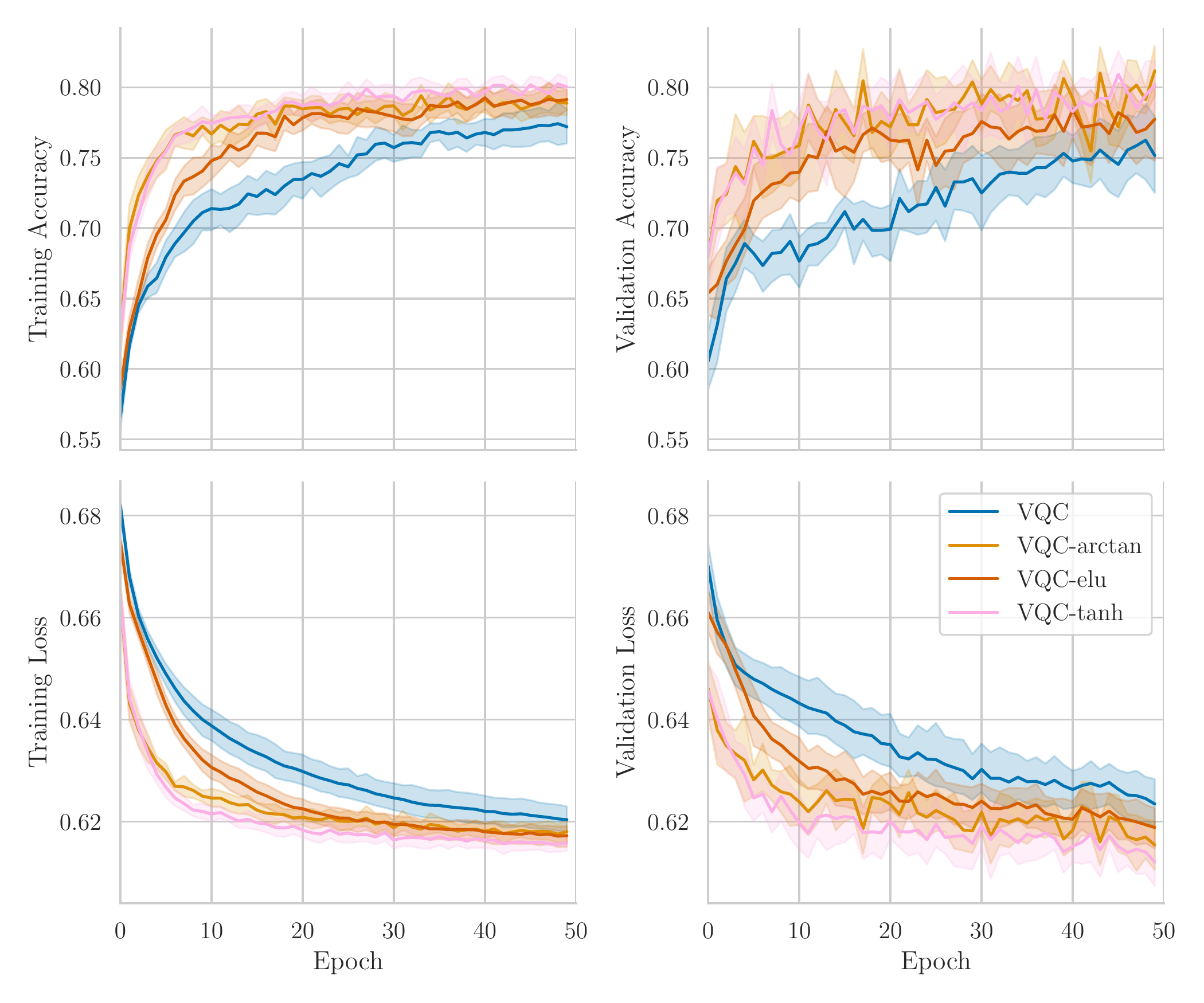}\label{fig:amp-heart}}
\subfloat[][Seeds Dataset]{\includegraphics[width=0.5\textwidth]{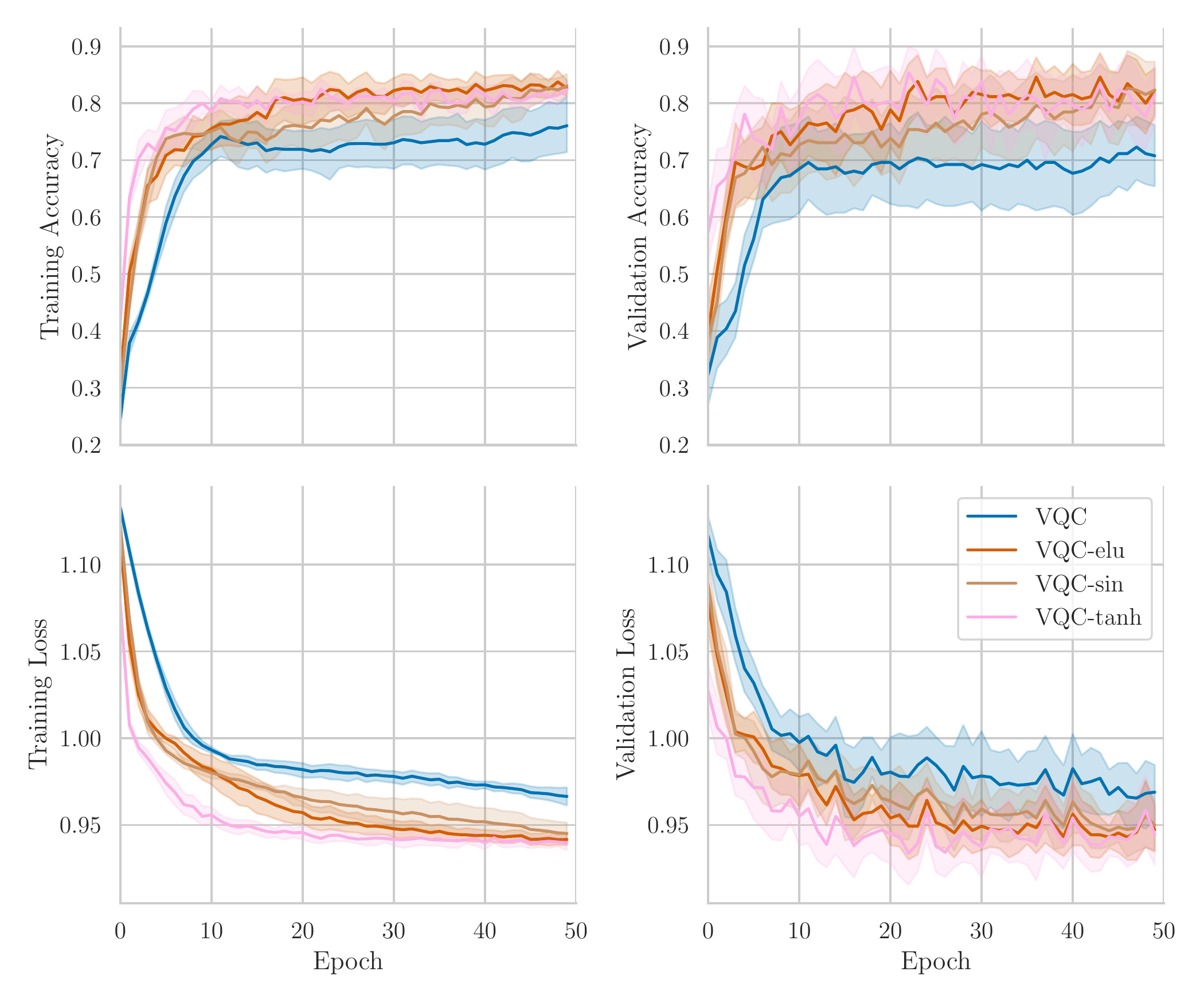}\label{fig:amp-seeds}}
\caption{Training and validation curves of the top 3 approaches for datasets Iris, Wine, Heart Desease and Seeds with Amplitude Embedding. In each epoch the algorithm processes $75\%$ of the total samples of each dataset for training.}
\label{fig:amp-results-valid}
\end{figure}

\begin{figure}[htb!]
 \centering
 \subfloat[][Iris Dataset]{\includegraphics[width=0.5\textwidth]{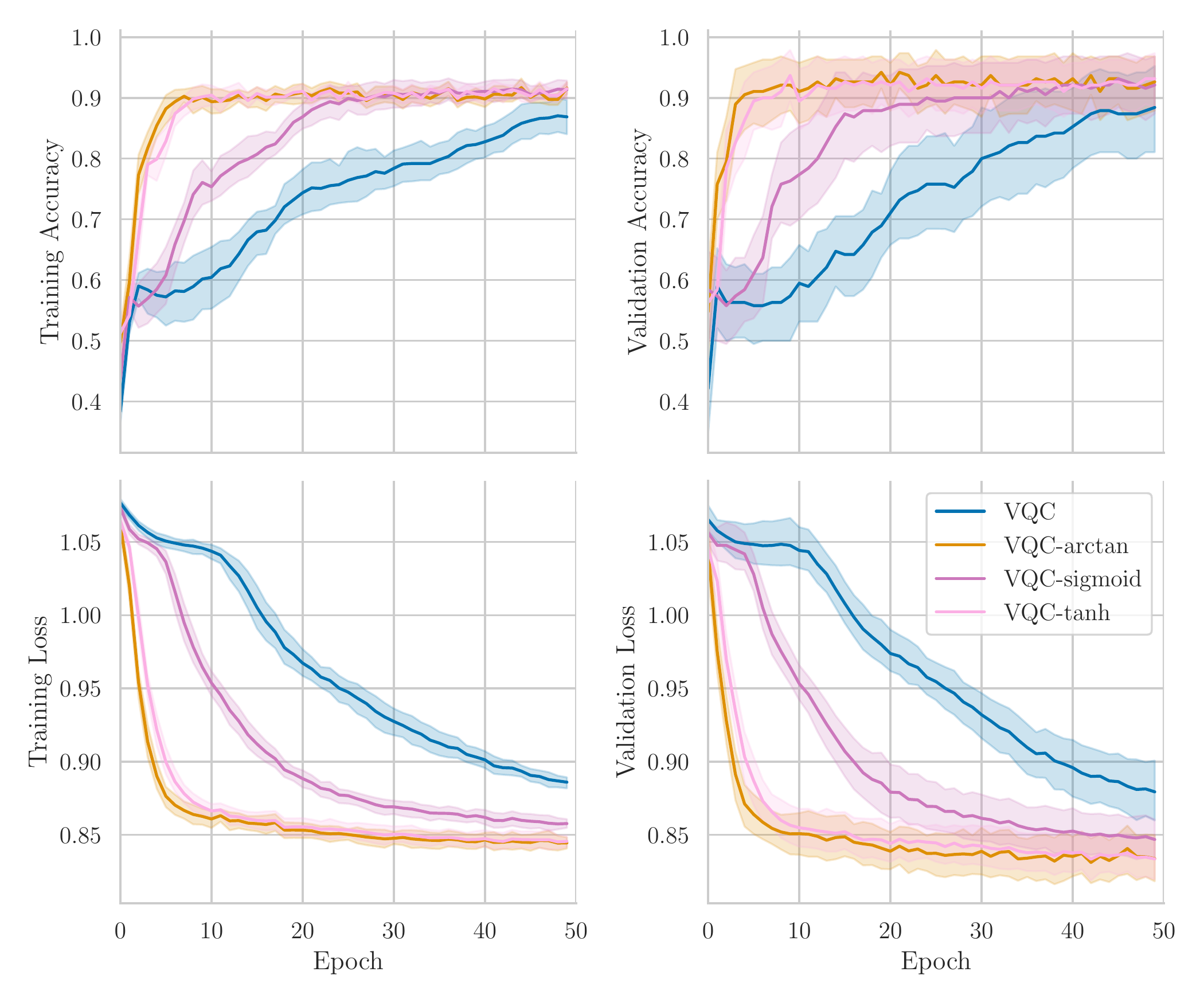}\label{fig:ang-iris}}
 \subfloat[][Abalone Dataset]{\includegraphics[width=0.5\textwidth]{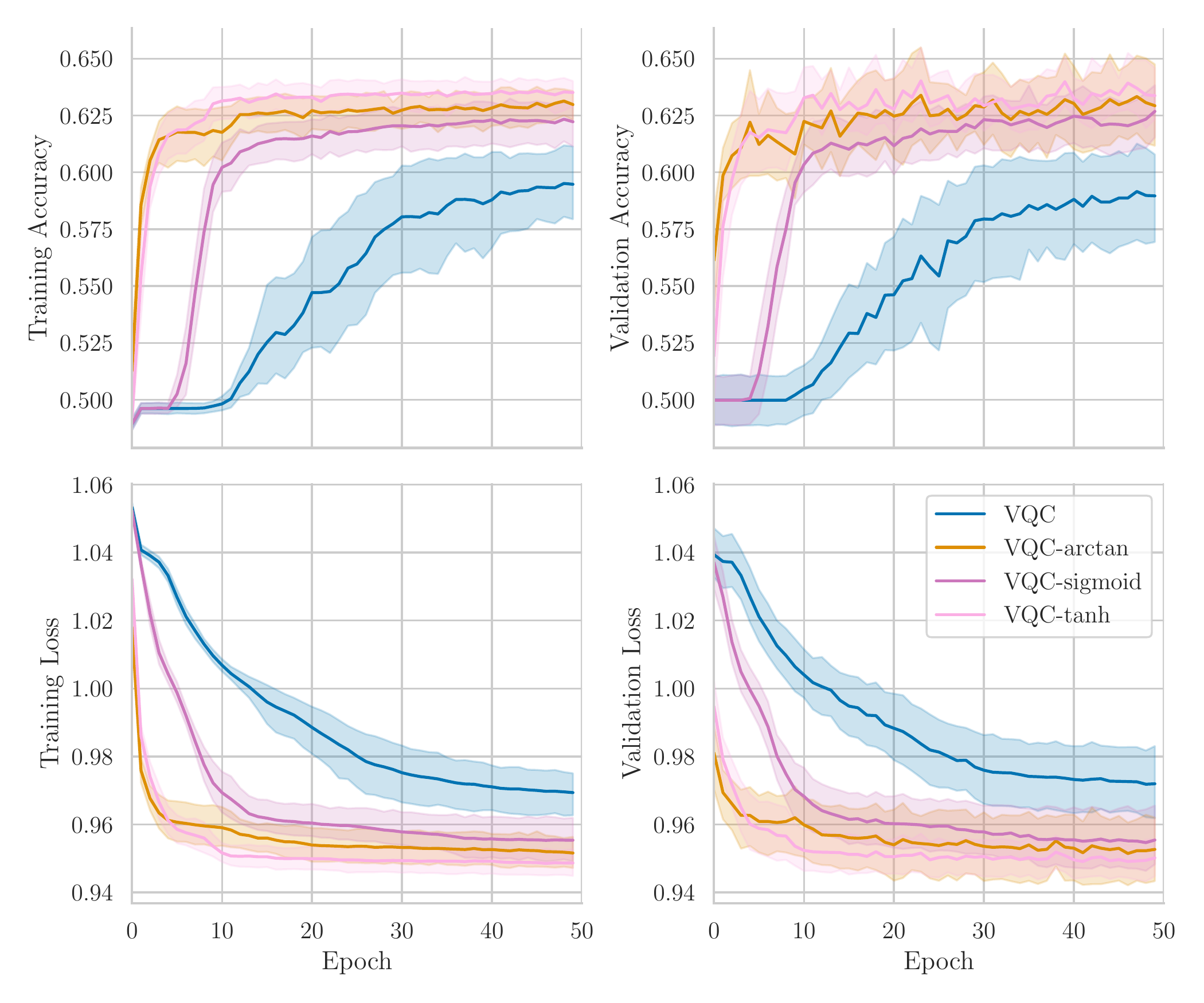}\label{fig:ang-abalone}}\\
 \subfloat[][Indian Diabetes Dataset]{\includegraphics[width=0.5\textwidth]{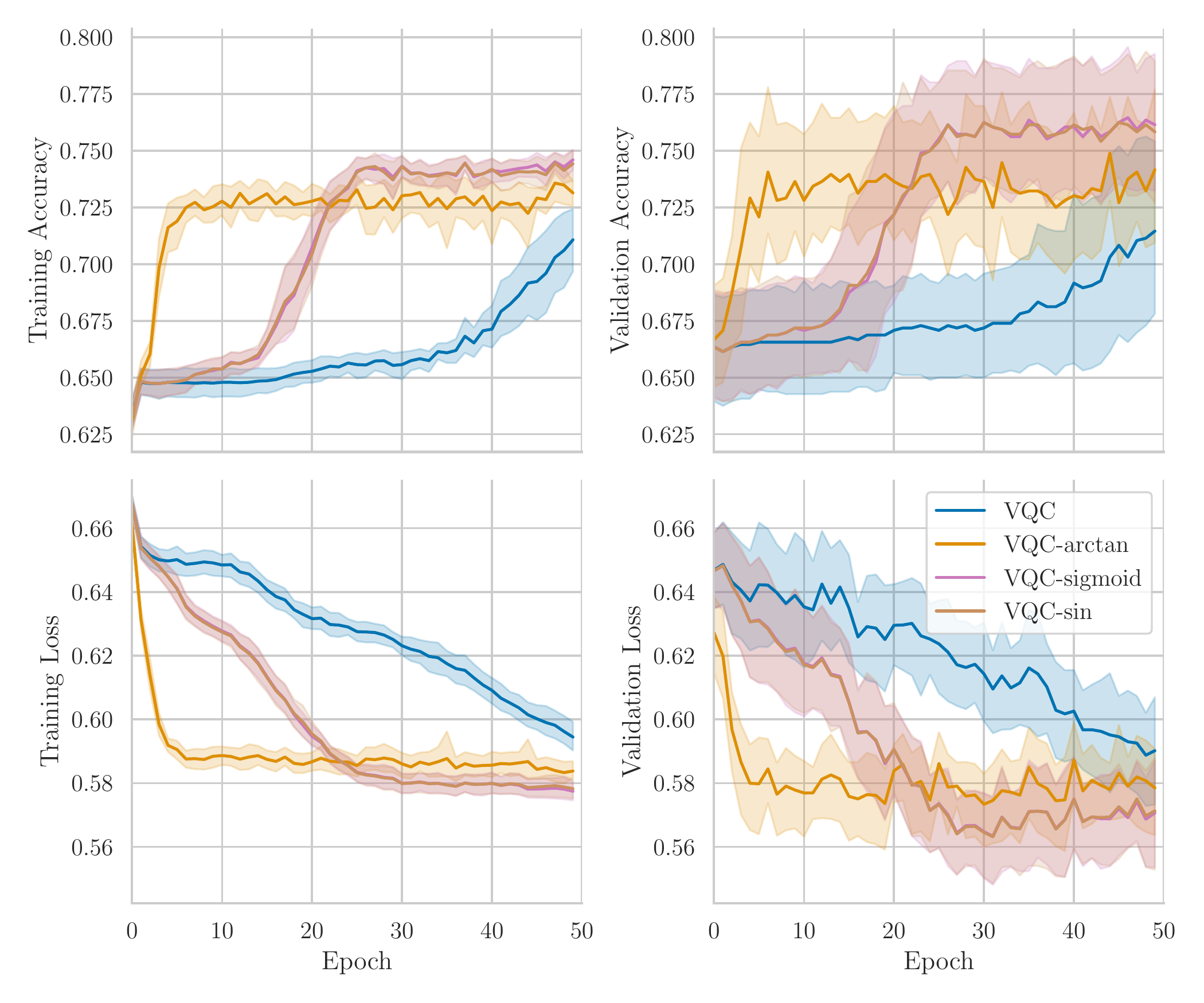}\label{fig:ang-diabetes}}
 \subfloat[][Seeds Dataset]{\includegraphics[width=0.5\textwidth]{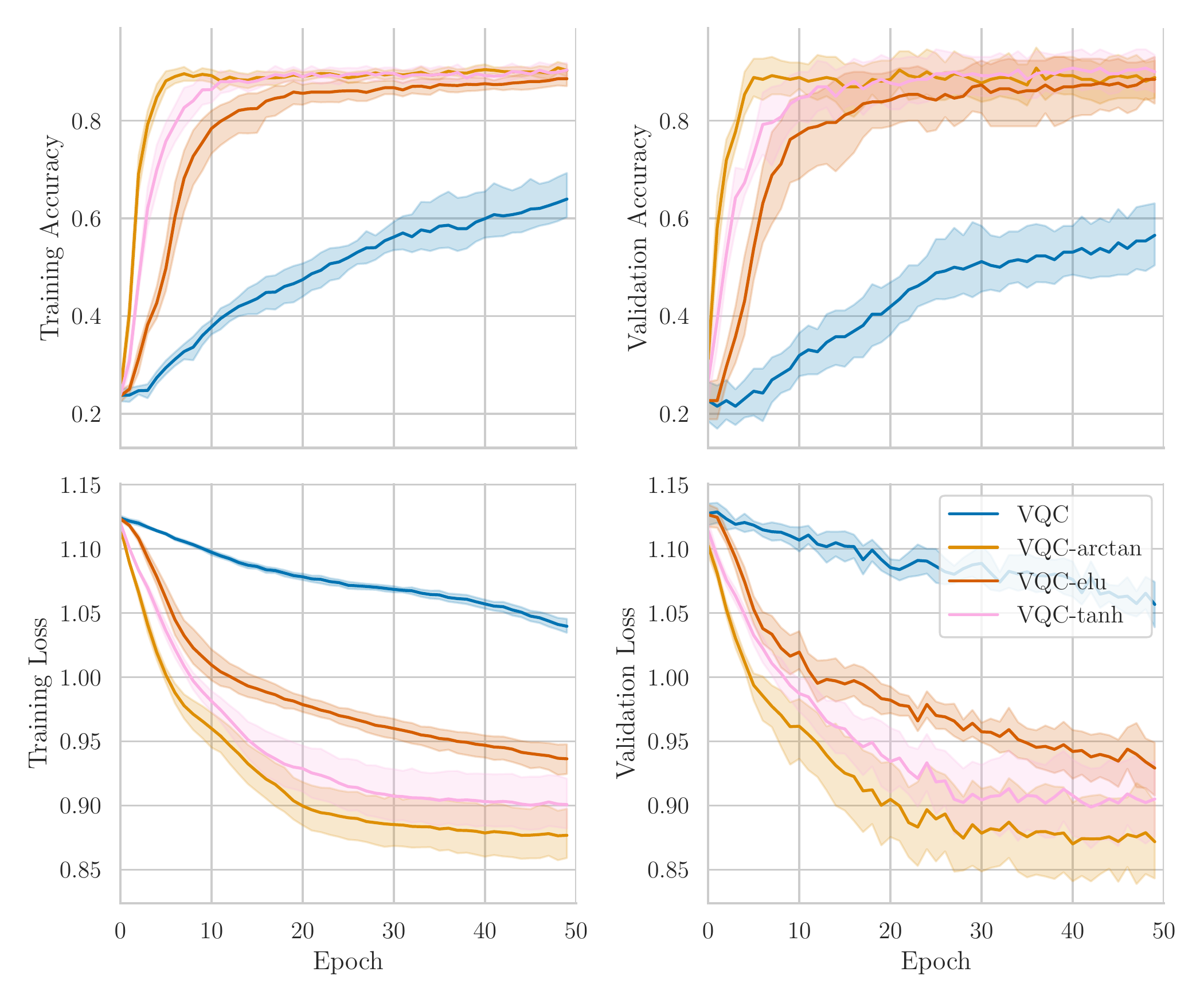}\label{fig:ang-seeds}}
 \caption{Training and validation curves of the top 3 approaches for datasets Iris, Abalone, Diabetis and Seeds with Angle Encoding. In each epoch the algorithm processes $75\%$ of the total samples of each dataset for training.}
 \label{fig:ang-results-valid}
\end{figure}

In this section we show and comment more in details, for each type of embedding, four of the eight datasets. For each dataset the top three performing re-mapping functions are shown. The complete results can be seen in the Appendix in Figure~\ref{fig:angle-all} and Figure~\ref{fig:amp-all}.
The selected results for Amplitude and Angle embedding are shown respectively in Figure~\ref{fig:amp-results-valid} and Figure~\ref{fig:ang-results-valid}. It can be seen how in both cases the convergence when applying the appropriate weight re-mapping function is much faster with respect to the VQC without any function applied. 
In order to properly estimate the impact that each re-mapping function has on the convergence of the VQC we propose a metric consisting in the difference in percentage between the validation accuracy of the VQC without and with a re-mapping function, as defined in Section~\ref{metrics}.
This metric has been calculated for each re-mapping function and for each dataset. The results can be seen in Table~\ref{tab:amp-convergence} for the Amplitude Embedding and in Table~\ref{tab:ang-convergence} for the Angle Embedding. 
We can see that in both cases the VQC-clamp has a difference of 0 with respect to the baseline, meaning that their performance is exactly the same. This is due to the fact that the weights here initialized and updated all are in the $[-\pi, \pi ]$ range, where the Clamp function behaves as an identity function, thus not affecting the performance of the VQC. Excluding the Clamp function we can see that, in both tables, averaging on all the datasets all the values are negative, meaning that the accuracy of the baseline is always smaller than the one of each approach at its point of convergence. Also for each dataset, apart from some exceptions, it can be seen the negative percentage, resembling a higher accuracy. 
In the case of Angle Embedding, which is usually the most common encoding, it can be seen how averaging on all the datasets at least an improvement of 10\% in convergence is always guaranteed.

\subsection{Average performance of re-mapping functions}\label{sec:average-performance}

    \begin{figure}[htb]
         \centering
         \subfloat[][Angle Embedding]{\includegraphics[width=0.5\textwidth]{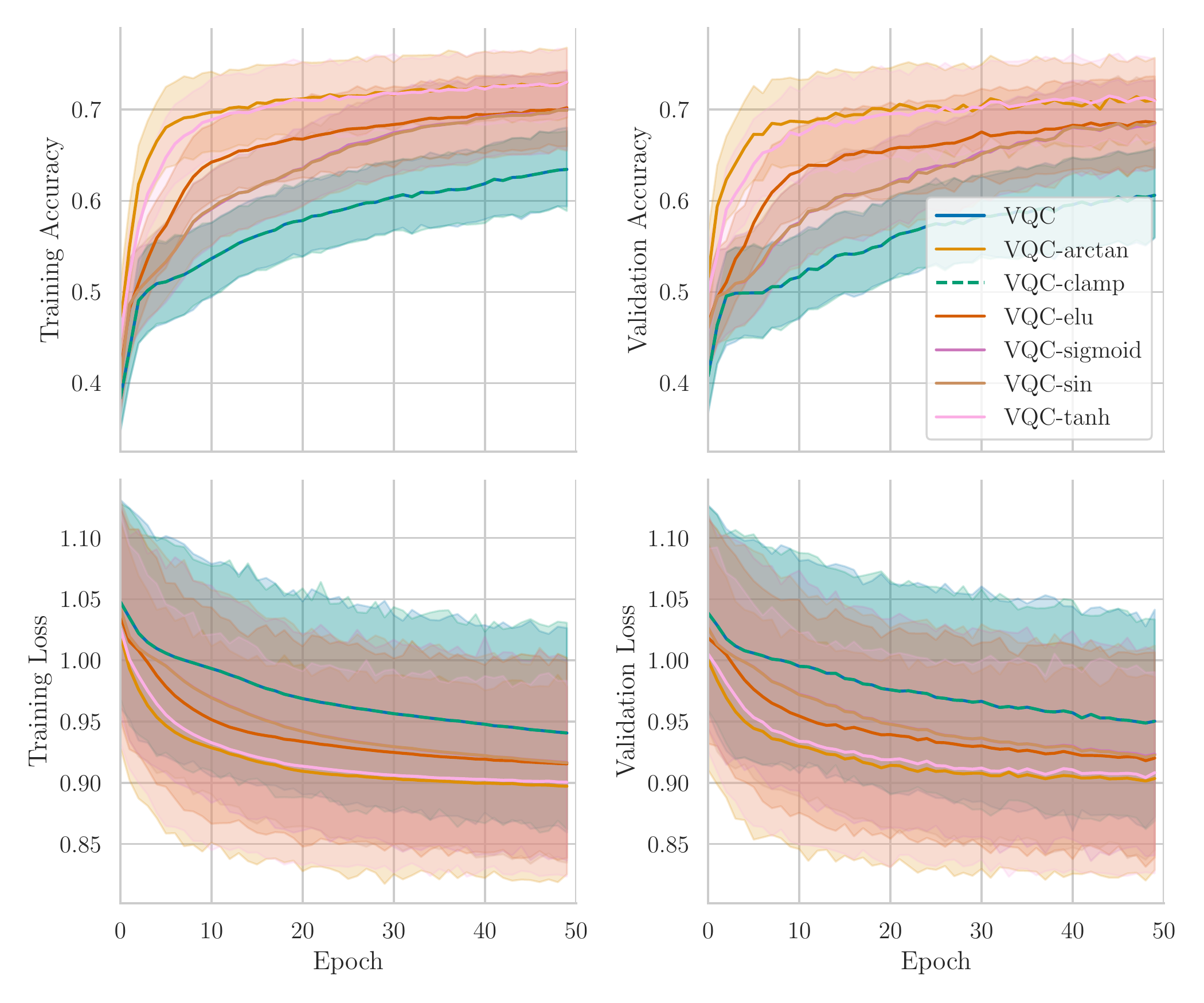}\label{fig:ang}}
         \subfloat[][Amplitude Embedding]{\includegraphics[width=0.5\textwidth]{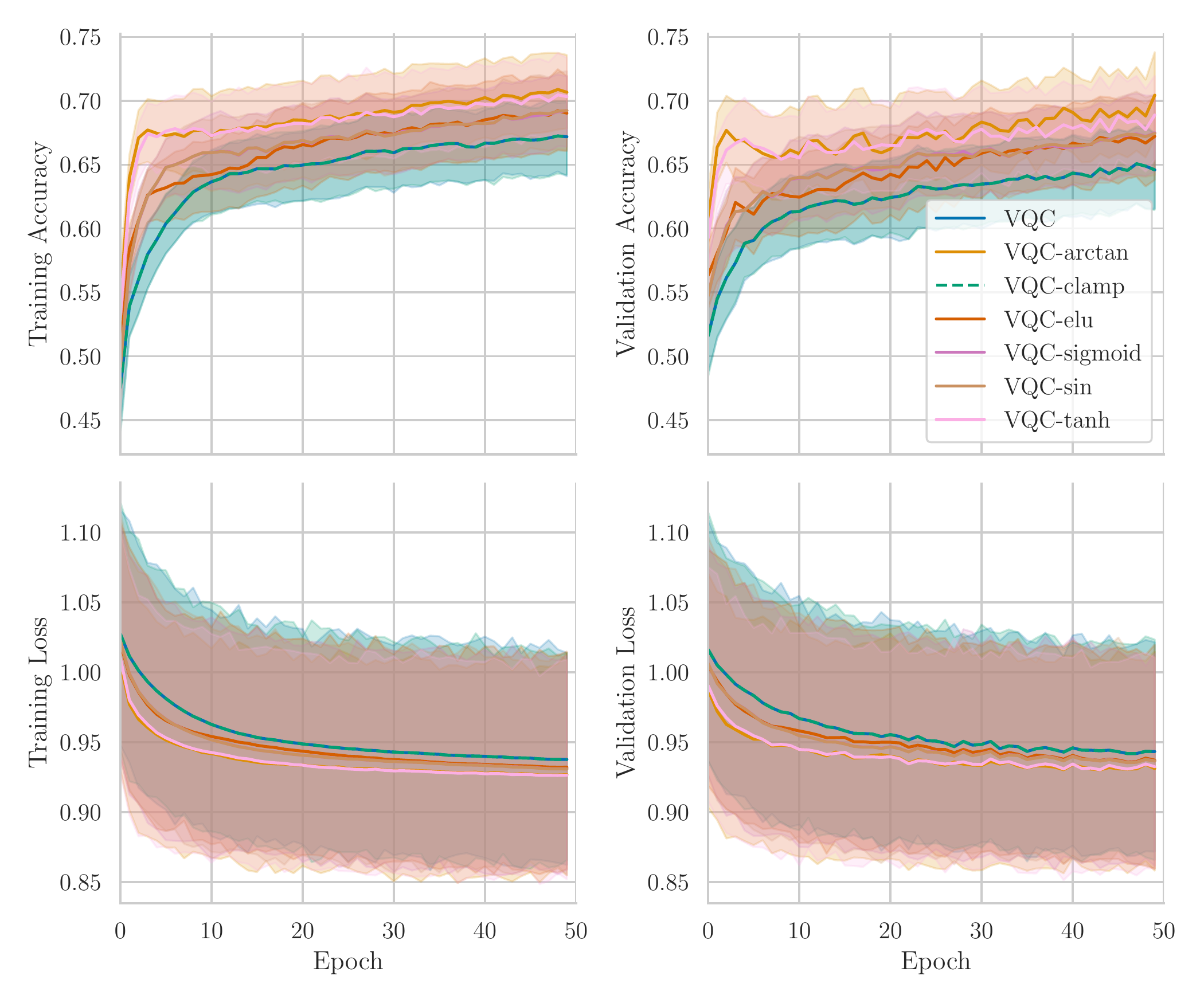}\label{fig:amp}}
         \caption{Training and validation curves for every weight re-mapping function averaged on all the datasets.}
         \label{fig:results-approaches}
    \end{figure}
    
    As a further analysis we studied the different re-mapping functions averaging the accuracy and loss on all the datasets. In Figure~\ref{fig:results-approaches} the results can be seen. Both for Angle and Amplitude embedding the faster convergence due to the re-mapping functions can be seen. With respect to the normal VQC, adding a re-mapping function increases the performances os the classifier.
    A more detailed insight is given in Table~\ref{tab:amp-convergence} and ~\ref{tab:ang-convergence}, in particular in the last row, where the improvement in performance has been averaged on all the datasets. It can be seen that, depending on the chosen Embedding, the re-mapping function that on average performs the best is different: in the case of Amplitude Embedding, the tanh function has the faster convergence, while in the case of Angle Embedding the best function is arctan. 
    Apart from these specific values which depend on the embedding and on the eight chosen datasets, the trend clearly indicates an improvement in the convergence speed of the classifier when constraining the weights of the VQC in their natural domain using an appropriate weight re-mapping function.

\subsection{Test Accuracy}\label{sec:test-accuracy}
    At the end of each training the final accuracy has been tested. In order to evaluate the impact on the final performance of the different weight re-mapping functions an ANOVA test of the test accuracies has been performed.
    
    \begin{table*}[htb!]
    \centering
    \begin{adjustbox}{width=\linewidth}
\begin{tabular}{llllllll}
\toprule
approach & VQC & VQC-arctan & VQC-clamp & VQC-elu & VQC-sigmoid & VQC-sin & VQC-tanh \\
dataset &  &  &  &  &  &  &  \\
\midrule
abalone & \pmb{$0.524 \pm 0.014$} & $0.520 \pm 0.014$ & \pmb{$0.524 \pm 0.014$} & $0.523 \pm 0.014$ & $0.522 \pm 0.014$ & $0.521 \pm 0.014$ & $0.522 \pm 0.014$ \\
banknote & \pmb{$0.701 \pm 0.022$} & $0.684 \pm 0.022$ & \pmb{$0.701 \pm 0.022$} & $0.698 \pm 0.022$ & $0.690 \pm 0.022$ & $0.690 \pm 0.022$ & $0.690 \pm 0.022$ \\
glass & $0.400 \pm 0.059$ & $0.437 \pm 0.060$ & $0.400 \pm 0.059$ & $0.411 \pm 0.059$ & $0.422 \pm 0.059$ & $0.422 \pm 0.059$ & \pmb{$0.441 \pm 0.060$} \\
heart\_deasease & $0.753 \pm 0.024$ & \pmb{$0.776 \pm 0.023$} & $0.753 \pm 0.024$ & $0.760 \pm 0.023$ & $0.759 \pm 0.023$ & $0.757 \pm 0.023$ & $0.767 \pm 0.023$ \\
indian\_diabetes & $0.655 \pm 0.030$ & $0.652 \pm 0.030$ & $0.655 \pm 0.030$ & \pmb{$0.657 \pm 0.030$} & $0.655 \pm 0.030$ & $0.655 \pm 0.030$ & $0.653 \pm 0.030$ \\
iris & $0.768 \pm 0.061$ & \pmb{$0.826 \pm 0.054$} & $0.768 \pm 0.061$ & $0.747 \pm 0.062$ & $0.732 \pm 0.064$ & $0.742 \pm 0.063$ & $0.821 \pm 0.055$ \\
seeds & $0.711 \pm 0.054$ & $0.778 \pm 0.050$ & $0.711 \pm 0.054$ & $0.789 \pm 0.049$ & $0.774 \pm 0.050$ & \pmb{$0.793 \pm 0.049$} & $0.781 \pm 0.050$ \\
wine & $0.643 \pm 0.062$ & \pmb{$0.848 \pm 0.047$} & $0.643 \pm 0.062$ & $0.726 \pm 0.058$ & $0.683 \pm 0.061$ & $0.678 \pm 0.061$ & $0.804 \pm 0.052$ \\
\midrule
$\varnothing$ & $0.644 \pm 0.041$ & \pmb{$0.690 \pm 0.037$} & $0.644 \pm 0.041$ & $0.664 \pm 0.040$ & $0.655 \pm 0.040$ & $0.657 \pm 0.040$ & $0.685 \pm 0.038$ \\
\bottomrule
\end{tabular}
    \end{adjustbox}
    \vspace*{3mm}
    \caption{Test Accuracy and 95\% confidence interval of tested mapping functions with Amplitude Embedding}
    \label{tab:test-acc-amp}
\end{table*}
    \begin{table*}[htb!]
    \centering
    \begin{adjustbox}{width=\linewidth}
\begin{tabular}{llllllll}
\toprule
approach & VQC & VQC-arctan & VQC-clamp & VQC-elu & VQC-sigmoid & VQC-sin & VQC-tanh \\
dataset &  &  &  &  &  &  &  \\
\midrule
abalone & $0.600 \pm 0.013$ & $0.635 \pm 0.013$ & $0.600 \pm 0.013$ & $0.624 \pm 0.013$ & $0.630 \pm 0.013$ & $0.628 \pm 0.013$ & \pmb{$0.639 \pm 0.013$} \\
banknote & $0.936 \pm 0.012$ & $0.949 \pm 0.010$ & $0.936 \pm 0.012$ & $0.946 \pm 0.011$ & $0.949 \pm 0.010$ & $0.949 \pm 0.010$ & \pmb{$0.952 \pm 0.010$} \\
glass & $0.315 \pm 0.056$ & \pmb{$0.459 \pm 0.060$} & $0.315 \pm 0.056$ & $0.426 \pm 0.059$ & $0.363 \pm 0.058$ & $0.363 \pm 0.058$ & \pmb{$0.459 \pm 0.060$} \\
heart\_deasease & $0.518 \pm 0.027$ & $0.677 \pm 0.026$ & $0.518 \pm 0.027$ & $0.625 \pm 0.026$ & $0.623 \pm 0.026$ & $0.622 \pm 0.026$ & \pmb{$0.691 \pm 0.025$} \\
indian\_diabetes & $0.711 \pm 0.029$ & $0.736 \pm 0.028$ & $0.711 \pm 0.029$ & $0.733 \pm 0.028$ & \pmb{$0.751 \pm 0.027$} & $0.747 \pm 0.028$ & $0.733 \pm 0.028$ \\
iris & $0.832 \pm 0.054$ & $0.884 \pm 0.046$ & $0.832 \pm 0.054$ & $0.858 \pm 0.050$ & $0.879 \pm 0.047$ & $0.879 \pm 0.047$ & \pmb{$0.900 \pm 0.043$} \\
seeds & $0.526 \pm 0.060$ & $0.893 \pm 0.037$ & $0.526 \pm 0.060$ & $0.863 \pm 0.041$ & $0.852 \pm 0.043$ & $0.856 \pm 0.042$ & \pmb{$0.904 \pm 0.035$} \\
wine & $0.422 \pm 0.064$ & \pmb{$0.517 \pm 0.065$} & $0.422 \pm 0.064$ & $0.404 \pm 0.064$ & $0.430 \pm 0.064$ & $0.435 \pm 0.065$ & $0.509 \pm 0.065$ \\
\midrule
$\varnothing$ & $0.607 \pm 0.039$ & $0.719 \pm 0.036$ & $0.607 \pm 0.039$ & $0.685 \pm 0.037$ & $0.685 \pm 0.036$ & $0.685 \pm 0.036$ & \pmb{$0.723 \pm 0.035$} \\
\bottomrule
\end{tabular}
    \end{adjustbox}
    \vspace*{3mm}
    \caption{Test Accuracy and 95\% confidence interval of tested mapping functions with Angle Embedding}
    \label{tab:test-acc-angle}
\end{table*}
    
    \noindent\textbf{Angle Embedding:} The ANOVA test has been first carried out averaging first across all the datasets. The results show a statistical difference and significance in the approaches, since we get $F(6,63) = 18.440, < = 0.001$. The performance of the approaches was further analyzed on individual datasets. On most of the datasets there were no significant differences in performance among the approaches: for Iris we get $F(6,63) = 0.534, p = 0.780$, for the Pina Indian Diabetes $F(6,63) = 1.186, p = 0.325$, for the Banknote Authentication Dataset $F(6,63) = 1.205, p = 0.315$, for the Wine Dataset $F(6,63) = 1.474, p = 0.202$. Statistical significance in the impact of the different approaches is instead obtained with the Abalone Dataset, with $F(6,63) = 3.138, p = 0.009$, glass dataset, $F(6,63) = 3.362, p = 0.006$, Heart Deseases, where $F(6,63) = 23.078, p = < 0.00$ and the Seeds dataset, where $F(6,63) = 35.402, p < 0.001$ \\
    
    \noindent\textbf{Amplitude Embedding:} The ANOVA revealed instead no significant differences between the final performance of the approaches using Amplitude Embedding, $F(6, 63) = 0.70$, $p = 0.65$. The performance of the approaches was also here further analyzed on individual datasets. On the abalone dataset, there were no significant differences in performance among the approaches, $F(6, 63) = 0.056$, $p = 0.999$. Similar results were observed for the banknote dataset, $F(6, 63) = 0.38$, $p = 0.89$, the glass dataset, $F(6, 63) = 0.31$, $p = 0.93$, the heart disease dataset, $F(6, 63) = 0.32$, $p = 0.92$, and the Indian diabetes dataset, $F(6, 63) = 0.016$, $p = 1.00$. On the iris dataset, there was a trend towards a difference in performance, but this did not reach statistical significance, $F(6, 63) = 1.55$, $p = 0.18$. The results were similar for the seeds dataset, $F(6, 63) = 1.17$, $p = 0.34$. However, on the wine dataset, there was a significant difference in performance between the approaches, $F(6, 63) = 4.01$, $p = 0.002$. \\

    \noindent These results suggest that adding an appropriate weight re-mapping function may not only improve the convergence speed of the classifier, as previously discussed, but also has an impact on its final accuracy. In our experiments this improvement in the final accuracy is significant when using the Angle Embedding as an encoding method, while not so much when using the Amplitude Embedding. 
    However, our results also highlight the importance of dataset characteristics in determining the performance of the different approaches, since it can be seen, with both types of embeddings, that the statistical significance varies on each dataset. Al already shown in the previous section, and here once again confirmed, the impact of the re-mapping functions depends on the chosen set of data. -In light of these findings, researchers and practitioners might need to consider both the properties of their dataset and try different re-mapping functions to improve their performance. Further studies are needed to understand the dataset features and the techniques that could predict the relative performance of the different approaches.
    Once again we highlight the fact that these ANOVA tests refer to the \textit{final} accuracy performance. Even though is some cases, like with Amplitude Embedding, this does not show a major improvement, the main result of introducing re-mapping functions, namely speeding up the convergence, still holds, as discussed in Section~\ref{sec:convergence-results}.

\subsection{Comparison to classical neural network}\label{sec:classical-comparison}

    \begin{figure}[htb]
        \centering
        \includegraphics[width=\textwidth]{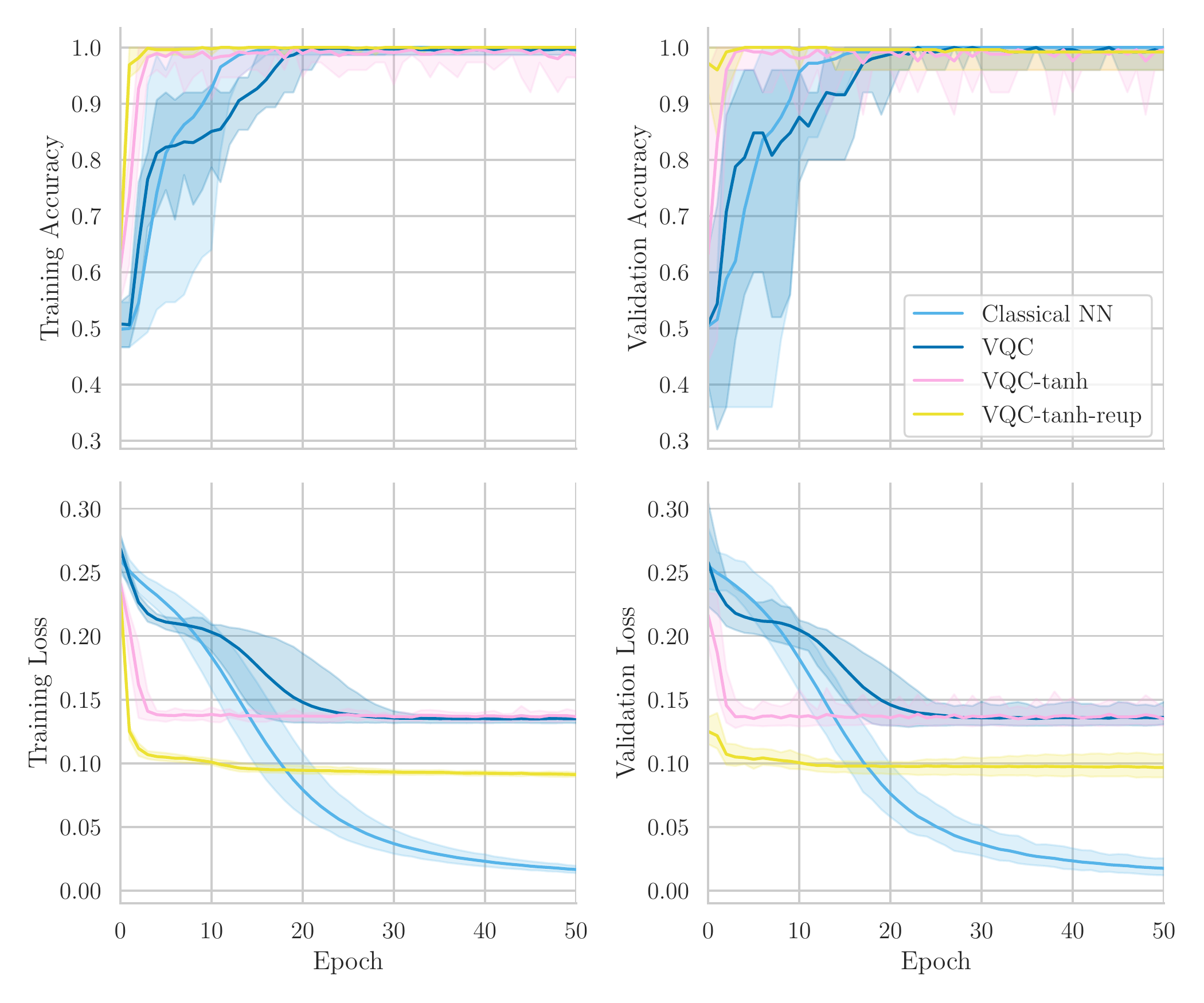}
        \caption{Comparison of the VQC without weight re-mapping, VQC-tanh, VQC-tanh with data re-uploading and a classical neural network, evaluated on the two-class iris dataset}
        \label{fig:classical-comp}
    \end{figure}
    
    \noindent One of the recurring questions when dealing with VQCs is how their performances compare to classical neural networks with a similar amount of parameters. In this section we thus compare with a small case study a VQC, a VQC-tanh, so with a weight re-mapping function, a classical neural network and a VQC-tanh with data re-uploading. More precisely, we trained the VQC on the 2-D Iris dataset, a reduced version of the Iris dataset already described where only two classes are employed instead of three. Since two qubits have been used to encode the information with Amplitude Embedding and the depth is of six layers, we get a total of 36 parameters. In order to give a classical neural network the same amount of parameters, we introduced only one hidden layers of 6 neurons. \\

    \noindent We can see the results in Figure~\ref{fig:classical-comp}. It's easy to see how the classical neural network and the VQC roughly have the same validation accuracy, even though the first converges slightly faster. Once the VQC is equipped with a re-mapping function though, the convergence once again increases dramatically, with the data re-uploading architecture performing slightly better than the other. This can also be seen more clearly in through the Validation Loss, where the most striking result is how the classical neural network reaches much lower losses than the quantum approximators. This results in a much smaller variance in the accuracy at the point of convergence, which instead does not shrink in the quantum case.

\section{\uppercase{Conclusion}}
\label{sec:conclusion}

In this paper, we build upon the work of \cite{koelle2023}, studying weight re-mapping functions in greater detail. This approach consists of re-scaling the weights of a variational quantum circuit so that they belong to their natural rotational domain, namely the $[-\pi, \pi]$ interval. We expand the scope of the work by integrating additional re-mapping functions into the VQC. Moreover, we study the performance of each function used in a classifier on eight different datasets, in contrast to the initial two proposed in the original work.  
Our key contribution lies in demonstrating that the weight re-mapping approach consistently accelerates the convergence speed. This convergence speed varies depending on the dataset, but we also tested the impact of each function by averaging across all datasets. It can be seen that functions that are steeper around the initialization point, like arctan or tanh, have a higher speedup in convergence. This is a particularly important result in the field of Quantum Machine Learning. In all practical applications of the NISQ era, the resources needed to train or evaluate a VQC are still very costly, and enhancing the convergence of a quantum circuit translates into substantial energy savings. We further demonstrate that, in certain cases, weight re-mapping can notably enhance test accuracy.
An intriguing question emerging from this work is: which dataset features influence the selection of the re-mapping function? As already mentioned, depending on the dataset structure, certain functions may be more effective than others. Further studies should be conducted to understand how the underlying properties of a dataset relate to the choice of the re-mapping function. 
As an alternative or complementary approach, one could examine various optimizers that factor in the periodicity of the weights in the rotational realm. 
Ultimately, this approach should be applied to other traditional machine learning tasks—such as Quantum Reinforcement Learning—to assess its efficacy in addressing more generalized problems.

\section*{\uppercase{Acknowledgements}}
Datsets used were taken from the UCI Machine Learning Repository \cite{Dua:2019}. This paper was partially funded by the German Federal Ministry of Education and Research through the funding program ``quantum technologies --- from basic research to market'' (contract number: 13N16196).

\clearpage
%
\bibliographystyle{splncs04}
\bibliography{main}

\begin{thebibliography}{10}
\providecommand{\url}[1]{\texttt{#1}}
\providecommand{\urlprefix}{URL }
\providecommand{\doi}[1]{https://doi.org/#1}

\bibitem{bellman1966dynamic}
Bellman, R.: Dynamic programming. Science  \textbf{153}(3731),  34--37 (1966)

\bibitem{benedetti_parameterized_2019}
Benedetti, M., Lloyd, E., Sack, S., Fiorentini, M.: Parameterized quantum
  circuits as machine learning models. Quantum Science and Technology
  \textbf{4}(4),  043001 (nov 2019). \doi{10.1088/2058-9565/ab4eb5},
  \url{https://dx.doi.org/10.1088/2058-9565/ab4eb5}

\bibitem{seeds}
Charytanowicz, M., Niewczas, J., Kulczycki, P., Kowalski, P.A., Łukasik, S.,
  Żak, S.: Complete Gradient Clustering Algorithm for Features Analysis of
  X-Ray Images, vol.~69, pp. 15--24 (01 2010). \doi{10.1007/978-3-642-13105-92}

\bibitem{heart}
Detrano, R.: International application of a new probability algorithm for the
  diagnosis of coronary artery disease. American Journal of Cardiology
  \textbf{64},  304--310 (1989)

\bibitem{Du_2020}
Du, Y., Hsieh, M.H., Liu, T., Tao, D.: Expressive power of parametrized quantum
  circuits. Physical Review Research  \textbf{2}(3) (2020).
  \doi{10.1103/physrevresearch.2.033125},
  \url{https://doi.org/10.1103\%2Fphysrevresearch.2.033125}

\bibitem{Dua:2019}
Dua, D., Graff, C.: {UCI} machine learning repository (2017),
  \url{http://archive.ics.uci.edu/ml}

\bibitem{fisher1936use}
Fisher, R.A.: The use of multiple measurements in taxonomic problems. Annals of
  eugenics  \textbf{7}(2),  179--188 (1936)

\bibitem{Harrow_2009}
Harrow, A.W., Hassidim, A., Lloyd, S.: Quantum algorithm for linear systems of
  equations. Physical Review Letters  \textbf{103}(15) (oct 2009).
  \doi{10.1103/physrevlett.103.150502},
  \url{https://doi.org/10.1103%2Fphysrevlett.103.150502}

\bibitem{koelle2023}
K{\"{o}}lle, M., Giovagnoli, A., Stein, J., Mansky, M.B., Hager, J.,
  Linnhoff{-}Popien, C.: Improving convergence for quantum variational
  classifiers using weight re-mapping. In: Rocha, A.P., Steels, L., van~den
  Herik, H.J. (eds.) Proceedings of the 15th International Conference on Agents
  and Artificial Intelligence, {ICAART} 2023, Volume 2, Lisbon, Portugal,
  February 22-24, 2023. pp. 251--258. {SCITEPRESS} (2023).
  \doi{10.5220/0011696300003393},
  \url{https://doi.org/10.5220/0011696300003393}

\bibitem{lloyd_quantum_2020}
Lloyd, S., Schuld, M., Ijaz, A., Izaac, J., Killoran, N.: Quantum embeddings
  for machine learning (Feb 2020), \url{http://arxiv.org/abs/2001.03622},
  arXiv:2001.03622 [quant-ph]

\bibitem{mitarai_quantum_2018}
Mitarai, K., Negoro, M., Kitagawa, M., Fujii, K.: Quantum circuit learning.
  Phys. Rev. A  \textbf{98},  032309 (Sep 2018).
  \doi{10.1103/PhysRevA.98.032309},
  \url{https://link.aps.org/doi/10.1103/PhysRevA.98.032309}

\bibitem{mottonen2004transformation}
Mottonen, M., Vartiainen, J.J., Bergholm, V., Salomaa, M.M.: Transformation of
  quantum states using uniformly controlled rotations (2004)

\bibitem{abalone_dataset}
Nash, W.J., Sellers, T.L., Talbot, S.R., Cawthorn, A.J., Ford, W.B.: The
  population biology of abalone (haliotis species) in tasmania. i. blacklip
  abalone (h. rubra) from the north coast and islands of bass strait. Sea
  Fisheries Division, Technical Report  \textbf{48}, ~p411 (1994)

\bibitem{nielsen_chuang_2010}
Nielsen, M.A., Chuang, I.L.: Quantum Computation and Quantum Information: 10th
  Anniversary Edition. Cambridge University Press (2010).
  \doi{10.1017/CBO9780511976667}

\bibitem{P_rez_Salinas_2020}
P{\'{e}}rez-Salinas, A., Cervera-Lierta, A., Gil-Fuster, E., Latorre, J.I.:
  Data re-uploading for a universal quantum classifier. Quantum  \textbf{4},
  ~226 (feb 2020). \doi{10.22331/q-2020-02-06-226},
  \url{https://doi.org/10.22331%2Fq-2020-02-06-226}

\bibitem{Schuld_2020}
Schuld, M., Bocharov, A., Svore, K.M., Wiebe, N.: Circuit-centric quantum
  classifiers. Physical Review A  \textbf{101}(3) (mar 2020).
  \doi{10.1103/physreva.101.032308},
  \url{https://doi.org/10.1103%2Fphysreva.101.032308}

\bibitem{Singh2019}
Singh, D., Singh, B.: Investigating the impact of data normalization on
  classification performance. Applied Soft Computing p. 105524 (05 2019).
  \doi{10.1016/j.asoc.2019.105524}

\bibitem{diabetes}
Smith, J.W., Everhart, J.E., Dickson, W., Knowler, W.C., Johannes, R.S.: Using
  the adap learning algorithm to forecast the onset of diabetes mellitus. In:
  Proceedings of the annual symposium on computer application in medical care.
  p.~261. American Medical Informatics Association (1988)

\bibitem{stoudenmire_supervised_2016}
Stoudenmire, E., Schwab, D.J.: Supervised {Learning} with {Tensor} {Networks}.
  In: Lee, D., Sugiyama, M., Luxburg, U., Guyon, I., Garnett, R. (eds.)
  Advances in {Neural} {Information} {Processing} {Systems}. vol.~29. Curran
  Associates, Inc. (2016),
  \url{https://proceedings.neurips.cc/paper/2016/file/5314b9674c86e3f9d1ba25ef9bb32895-Paper.pdf}

\bibitem{parvus}
Vandeginste, B.: Parvus: An extendable package of programs for data
  exploration, classification and correlation, m. forina, r. leardi, c.
  armanino and s. lanteri, elsevier, amsterdam, 1988, price: Us \$645 isbn
  0-444-43012-1. Journal of Chemometrics  \textbf{4}(2),  191--193 (1990).
  \doi{https://doi.org/10.1002/cem.1180040210},
  \url{https://analyticalsciencejournals.onlinelibrary.wiley.com/doi/abs/10.1002/cem.1180040210}

\bibitem{wilson_quantum_2019}
Wilson, C.M., Otterbach, J.S., Tezak, N., Smith, R.S., Polloreno, A.M.,
  Karalekas, P.J., Heidel, S., Alam, M.S., Crooks, G.E., da~Silva, M.P.:
  Quantum {Kitchen} {Sinks}: {An} algorithm for machine learning on near-term
  quantum computers (Nov 2019). \doi{10.48550/arXiv.1806.08321},
  \url{http://arxiv.org/abs/1806.08321}, arXiv:1806.08321 [quant-ph]

\end{thebibliography}

\section*{\uppercase{Appendix}}
\begin{figure}[htb]
     \centering
     \includegraphics[width=0.85\textwidth]{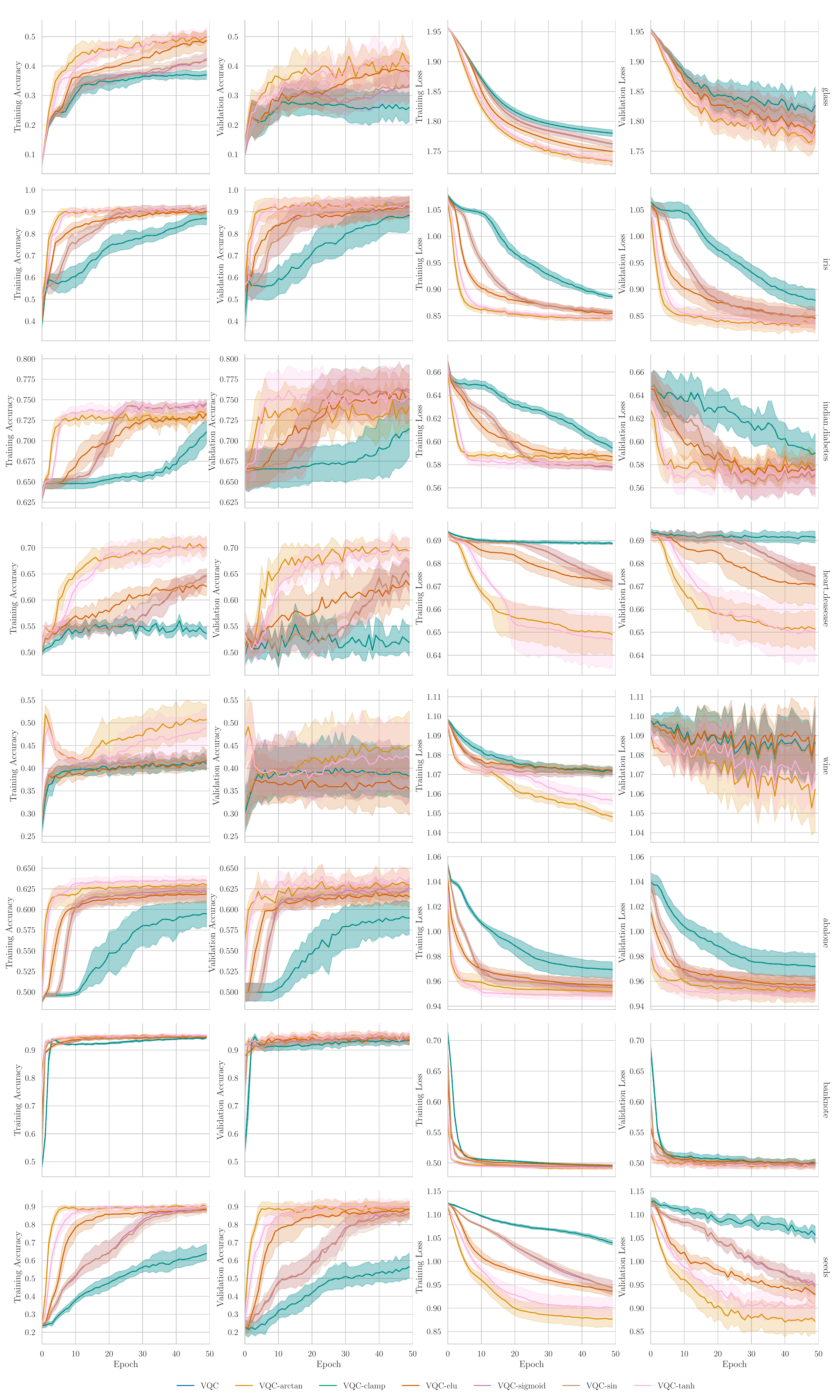}
     \caption{Training and validation curves for each dataset and approach using Angle Embedding (x-axis)} 
     \label{fig:angle-all}
\end{figure}

\begin{figure}[htb]
     \centering
     \includegraphics[width=0.85\textwidth]{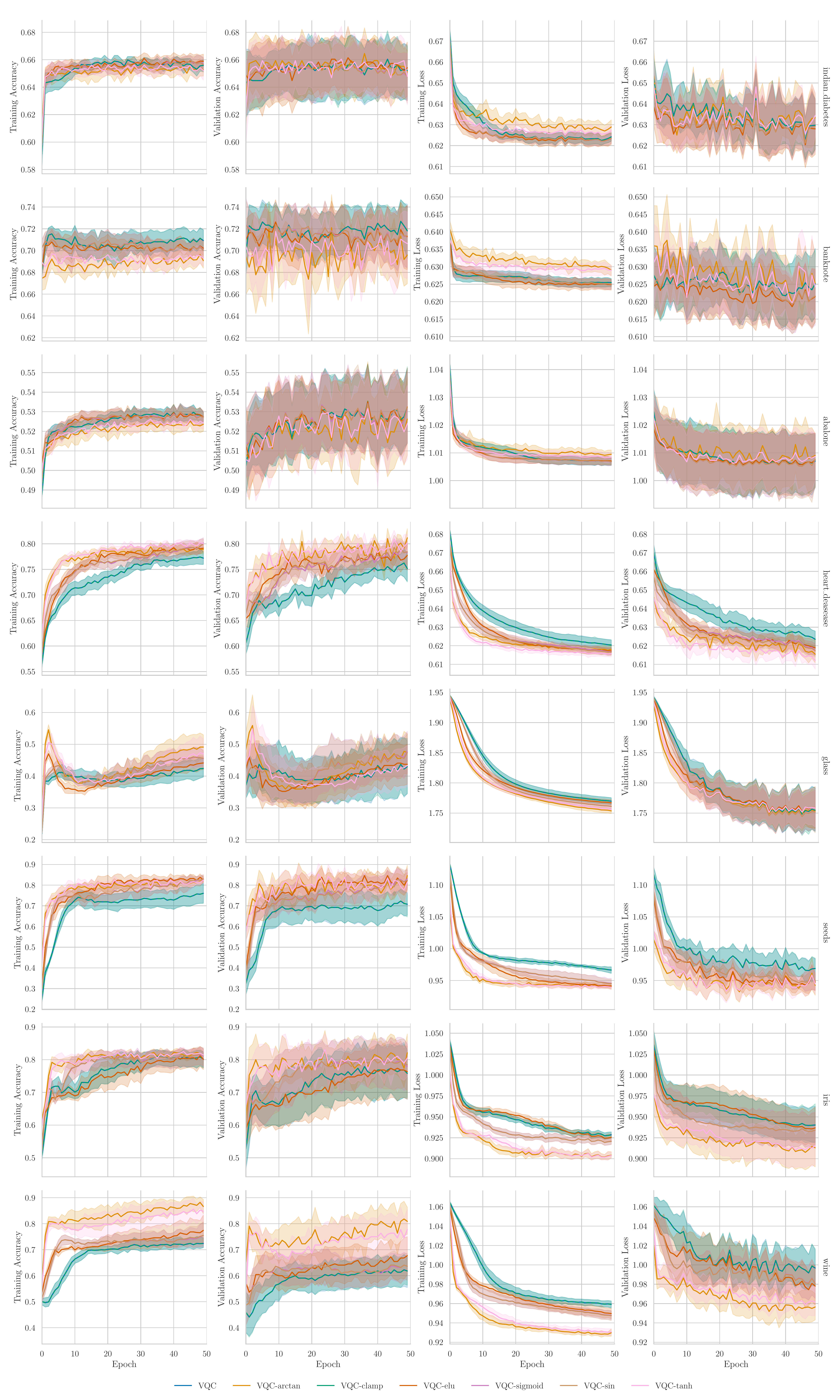}
     \caption{Training and validation curves for each dataset and approach using Amplitude Embedding}
     \label{fig:amp-all}
\end{figure}

In this section, we present the full evaluation results for all eight datasets and seven approaches. In Figure \ref{fig:angle-all}, we show the experiments for circuits with Angle Embedding (x-axis). Furthermore, the experiments for the Amplitude Embedding circuits can be found in Figure \ref{fig:amp-all}.

\end{document}